\documentclass[sigconf]{acmart}

\usepackage[capitalize]{cleveref}

\usepackage{tabularx}

\definecolor{changeNoteColor}{rgb}{0.1,0.6,1}
\newcommand{\changenote}[1]{#1} %

\AtBeginDocument{%
  \providecommand\BibTeX{{%
    \normalfont B\kern-0.5em{\scshape i\kern-0.25em b}\kern-0.8em\TeX}}}

\copyrightyear{2022} 
\acmYear{2022} 
\setcopyright{acmlicensed}\acmConference[UIST '22]{The 35th Annual ACM Symposium on User Interface Software and Technology}{October 29-November 2, 2022}{Bend, OR, USA}
\acmBooktitle{The 35th Annual ACM Symposium on User Interface Software and Technology (UIST '22), October 29-November 2, 2022, Bend, OR, USA}
\acmPrice{15.00}
\acmDOI{10.1145/3526113.3545672}
\acmISBN{978-1-4503-9320-1/22/10}

\definecolor{DanielsColor}{rgb}{0.9,0.6,0.1}

\definecolor{HaisColor}{rgb}{0.9,0.3,0.9}

\definecolor{FloriansColor}{rgb}{0,0.3,0.9}

\definecolor{KarimsColor}{rgb}{0.3,0.9,0.3}

\newcommand{\participant}[2]{$P_{#1}$, study #2}
\newcommand{\pquote}[3]{\textit{``#1''} (\participant{#2}{#3})}

\begin{document}

\title{Beyond Text Generation: Supporting Writers with Continuous Automatic Text Summaries}

\author{Hai Dang}
\orcid{0000-0003-3617-5657}
\email{hai.dang@uni-bayreuth.de}
\affiliation{%
  \institution{Department of Computer Science, University of Bayreuth}
  \city{Bayreuth}
  \country{Germany}
}

\author{Karim Benharrak}
\email{Karim.Benharrak@uni-bayreuth.de}
\affiliation{%
  \institution{Department of Computer Science, University of Bayreuth}
  \city{Bayreuth}
  \country{Germany}
}

\author{Florian Lehmann}
\orcid{0000-0003-0201-867X}
\email{florian.lehmann@uni-bayreuth.de}
\affiliation{%
  \institution{Department of Computer Science, University of Bayreuth}
  \city{Bayreuth}
  \country{Germany}
}

\author{Daniel Buschek}
\orcid{0000-0002-0013-715X}
\email{daniel.buschek@uni-bayreuth.de}
\affiliation{%
  \institution{Department of Computer Science, University of Bayreuth}
  \city{Bayreuth}
  \country{Germany}
}

\renewcommand{\shortauthors}{Dang et al.}

\begin{teaserfigure}
    \centering
    \includegraphics[width=0.95\textwidth]{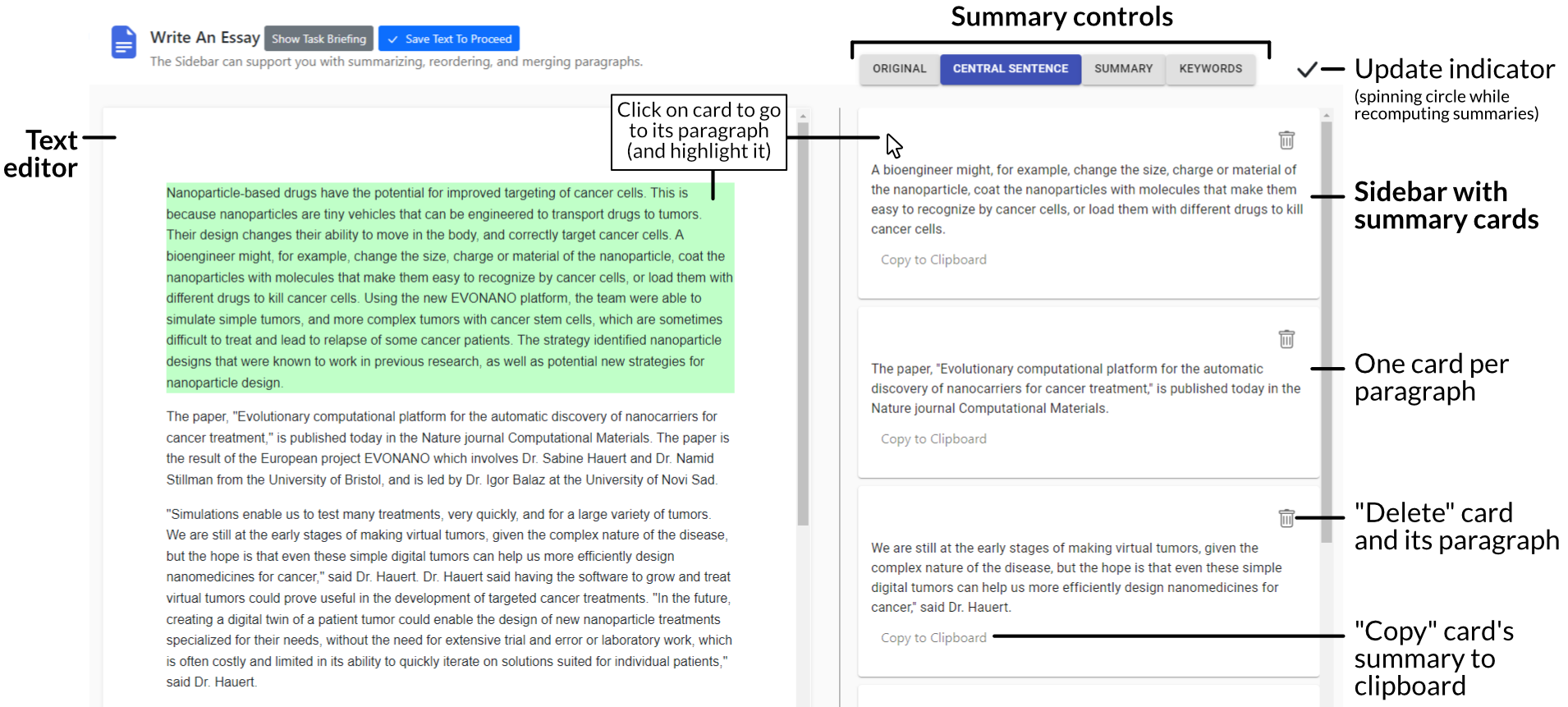}
    \caption{Our UI (in the final version for study 2) with text editor (left) and a sidebar (right), which shows one card per paragraph. Typing is possible only in the editor. The cards' content is controlled with the top buttons (original, central sentence, summary, keywords). These summaries are updated while typing. Editor and sidebar can be scrolled independently. Clicking a card scrolls to its paragraph in the editor and highlights it (in green for 1 second).}
    \label{fig:teaser}
    \Description{A screenshot of the interactive prototype. A paragraph is highlighted in green. To the right of the main text editor view there is a sidebar with three cards displaying a summarized version of the text paragraphs in the main editor.}
\end{teaserfigure}

\begin{abstract}
We propose a text editor to help users plan, structure and reflect on their writing process. It provides continuously updated paragraph-wise summaries as margin annotations, using automatic text summarization. Summary levels range from full text, to selected (central) sentences, down to a collection of keywords. To understand how users interact with this system during writing, we conducted two user studies (N=4 and N=8) in which people wrote analytic essays about a given topic and article. As a key finding, the summaries gave users an external perspective on their writing and helped them to revise the content and scope of their drafted paragraphs. People further used the tool to quickly gain an overview of the text and developed strategies to integrate insights from the automated summaries. More broadly, this work explores and highlights the value of designing AI tools for writers, with Natural Language Processing (NLP) capabilities that go beyond direct text generation and correction.
\end{abstract}




\maketitle

\section{Introduction}\label{sec:intro}

Writing is important and ubiquitous for many people. Numerous tools have been developed to aid writers in their creative process by automatically performing spell and grammar checking or suggesting continuation phrases to increase human performance and productivity. Yet, good writing goes beyond correct spelling and grammar -- it needs to both convey the writer's intention and address the reader's needs. Writing is a complex cognitive activity, as pointed out by %
\citet{flower1981cognitive}. In particular, writing interweaves various sub-activities. Revision is one of them, which is important but often challenging for several reasons: (1) \textit{Practically}, investing into preparations for writing or repeated revisions often stands at a tension to finishing the work. %
(2) \textit{Organizationally}, as the body of text grows, it is increasingly difficult to oversee the draft and maintain a big picture during writing. (3) \textit{Skill-related}, it is tempting, in particular for amateurs, to focus on local revisions (e.g. word-level) instead of scope and structure \cite{king_reverse_2012}. 

Despite these challenges, providing assistance for the specific activity of text revision has remained under-explored in HCI research, as pointed out in the community~\cite{Buschek2021chi, Vertanen2017, Vertanen2016}. Recent work here focuses on the aforementioned local revisions (e.g. word replacement~\cite{Han2020, Li2020}, error correction~\cite{Arif2016, Cui2020, Zhang2019}), increasingly applying Artificial Intelligence (AI) methods from Natural Language Processing (NLP). However, such co-creative use of AI has not been examined yet for higher-level revisions. %
These aspects motivate our research question for this paper:
\textit{How might we enable writers to benefit from computational (NLP/AI) capabilities beyond text generation and correction, in particular for revision?}

We address this question by exploring how we might support reflection and (structural) revision via \textit{automatic summarization}. This approach was inspired by taking an existing cognitive strategy for revision as a starting point, namely \textit{reverse outlining}~\cite{king_reverse_2012}: Writers manually create reverse outlines \textit{after} a part of the draft has been written, summarizing it, to reflect and identify opportunities for improvements. This guided our design and prototype: We automate this summarization to continuously update paragraph-wise summaries shown to writers next to their text. 

We tested this in two in-depth user studies with a total of 12 participants writing analytic essays about a given article. %
The summaries facilitated reflection and overview: They gave users an external perspective on their writing (e.g. comparing own expectations against what's in the summary). People used this to reflect on content and scope of paragraphs (e.g. identifying redundancy), leading to revisions. They further used the tool to quickly gain an overview of the given article, and developed strategies to integrate insights from the automated summaries into their text. 

In summary, we contribute a text editor prototype with automatic text summaries and its evaluation in two user studies. %
More broadly, our work contributes to the literature on interactive NLP, text interaction and co-creative AI: It explores and highlights the value of assisting writers with NLP capabilities beyond direct text generation and correction.

\section{Background and Related Work}\label{sec:related_work}
This work draws upon the areas of writing reseach, AI and NLP, and reading and writing support tools.

\subsection{What is a Summary and What is its Goal?}\label{sec:related_work_summaries}
Here we clarify our understanding of the term ``summary'' in this paper.
Overall, we align with the definition by \citet{radev_introduction_2002}, who define a summary ``as a text that is produced from one or more texts, that conveys important information in the original text(s), and that is no longer than half of the original text(s) and usually significantly less than that.'' %
They further state that the ``main goal of a summary is to present the main ideas in a document in less space'', which fits to our application as well, although we target the paragraph-level instead of the whole document. They also highlight a distinction in the literature between \textit{indicative} summaries (what is the text about, without specific content) and \textit{informative} ones (short version of the text). Our concept and system use the informative type.

Moreover, \citet{gelbukh_decomposition_2009} distinguish between \textit{objective} and \textit{interpretive} summaries to inform their work on a book summarisation system: The former type objectively summarises the plot, while the latter also integrates the subjective view of the person writing the summary. This classification is interesting to consider for our use case: Our system creates \textit{objective} (automatic) summaries of the user's text in order to prompt and facilitate the user's \textit{own interpretation} (i.e. self-reflection).

Finally, \textit{extracts} are short versions reusing parts of the full text verbatim, while \textit{abstracts} are short recreations~\cite{radev_introduction_2002}. Our concept and system explore both types (see \cref{sec:concept}).

\subsection{Automatic Text Summarization}
There are two main NLP approaches for automatically creating summaries of a given text~\cite{Kryscinski2019}: \textit{Extractive} methods aim to select the most important/relevant parts of the text. This selection directly yields the summary. In contrast, \textit{abstractive} methods aim to write a new piece of text to serve as a summary, similar to what most humans would do when tasked to summarize a text. There are also methods that mix aspects of both approaches, such as pointer concepts~\cite{See2017}, which allow a model to copy some parts verbatim while (re)writing others. In our prototype, we employ a \textit{TextRank}~\cite{Mihalcea2004textrank} approach for extractive summaries and the \textit{T5} model~\cite{Raffel2020} for abstractive summaries (details in \cref{sec:implementation}).

\subsection{Reading and Writing Support Systems}\label{sec:related_work_tools}

A number of interactive systems support reading and writing with summarization:
For example, \citet{Leiva2018} used extractive summarization to make websites more responsive to the device size -- not only by adapting the visual layout but also the length of its text content. 
The system by \citet{wang2021screen2words} summarizes mobile UIs to create succinct descriptions of screen content, and \citet{terHoeve2020} proposed a conversational UI (chatbot, voice assistant) that reports information from a document when asked about it. %
\citet{Li2021uist} combined speech recognition with summarization to show text snippets that help users navigate long audio content.

Related, but not using summarization, are the many tools that support creating text; this is only a small overview: Already in 1982, \citet{macdonald1982writer} built a ``writer's workbench'', which could comment on stylistic features. %
Recent work typically generate text suggestions (e.g. for the next sentence \cite{roemmele2015creative} or paragraph~\cite{yang_ai_2022} or in a sidebar~\cite{elephant_tochi2022}). More broadly, this co-creation through interleaved human writing and AI-generated text is a common approach (e.g. \cite{Buschek2021chi, Chen2019, swanson2008say, yuan_wordcraft_2022, lee_coauthor_2022}). Further work explored controllable generation of plots with AI \cite{Tambwekar2018controllable, chung_talebrush_2022}, or addressed writing poetry~\cite{Hafez2017}, metaphors~\cite{Gero2019chi}, slogans~\cite{clark2018}, fictional characters~\cite{Schmitt2021}, and emails~\cite{Buschek2021chi, Kannan2016}. \citet{Strobl2019} provide a recent survey of 44 academic writing tools. Further recent work supports revision on the level of word replacement~\cite{Han2020, Li2020} and (typing) error correction~\cite{Arif2016, Cui2020, Zhang2019}, partly using NLP models. Finally, \citet{arnold_generative_2021} recently highlighted opportunities for supporting writing without generating text, which motivates our design direction here.

In summary, the literature has mainly used (1) automatic summaries as reading support, and (2) text generation as writing support. We explore the remaining combination, namely \textit{using (AI) summarization to support writing} as it happens. %
Conceptually, we build on an existing writing strategy, described next (\cref{sec:related_work_reverse_outlining}). %

\subsection{Text Summarization in Writing Research, Practice and Instruction}\label{sec:related_work_reverse_outlining}

Summarizing text is an integral step in the strategy of \textit{reverse outlining}~\cite{king_reverse_2012} (or \textit{backward-outlining}~\cite{saltz_harvard_1998} or \textit{post-outlining}~\cite{uni_tennessee_writing_strategies_2007}), which has been called ``the Swiss army knife of revising''~\cite{duke_university_revising_2021}: Instead of a typical outline that is created before drafting a text, a reverse outline is created after (a version of) the draft has been written. Concretely, the writer summarizes each paragraph. If done on paper, the result can be cut into one note per paragraph for easy rearrangement~\cite{messuri_revision_2016, messuri_writing_2016}. These summaries support self-reflection about the text's structure (e.g. \textit{Is the text's claim supported by each part?}, \textit{Are the aspects covered in a suitable order?}, \textit{Do paragraphs contain a single thought each?}) and subsequent actions based on the gained insights (e.g. reordering, splitting, merging or deleting paragraphs). %
This strategy is a part of many university writing guides and courses (e.g.~\cite{duke_university_revising_2021, park_reverse_2008, saltz_harvard_1998, university_of_wisconsin-madison_reverse_2021}), and other teaching~\cite{king_reverse_2012, tully_reverse_2019, levan2017self}, and valued by professional writers (e.g.~\cite{hamburger_outlining_2013}). %

More broadly, reverse outlining is an instance of (self-)annotation. Annotations are ``concise descriptions'' of a work and can be descriptive or evaluative~\cite{yayli2012tracing}: Our use of AI summaries is descriptive since it does not include statements on what is good or bad about the summarized text. As suggested in studies with students~\cite{Cresswell2000}, the act of self monitoring afforded in this way can improve control over one's own writing. 
In this light, our work explores AI-supported self monitoring during writing. %

\section{Concept Development}\label{sec:concept}

Here we report on our concept development, starting with the core conceptual inspiration, before covering UI aspects and interaction. %

\subsection{Core Concept: AI Text Summarization as (Self-)Annotation
}

As described in \cref{sec:related_work_reverse_outlining}, reverse outlining is an effective strategy for reflecting on and revising a text structurally -- but without interactive support yet. In our concept development, we took this as a starting point to explore how we might support it technically in an interactive system. Concretely, reverse outlining offers two key conceptual aspects that we pick up on in our design: \textit{Annotating} what has been written, and the fact that this is \textit{one's own text} (and not that of another person). We comment on both in relation to our concept and design direction in more detail here:

\citet{levan2017self, yayli2012tracing} emphasize the value of self-assessment with self-annotation in structuring and understanding text during writing. They mention ``sideline commentary for their writing'' as a strategy for students to externalize their thoughts and gain an overview of their work in progress. As a key part of our conceptual exploration, we decided to automate the summarization step of such self-annotation with AI to support self-assessment. While the act of manual summarization can be seen as a part of the reflection process, we assume that it is also an initial hurdle to enter a reflection and revision step (e.g. investing time and effort into summarizing what you have already written vs writing more). By making summaries automatically available to writers our concept thus is intended to invite and support reflection -- and potentially revision, if identified as needed. 

Automatic summaries are not a ``zero cost'' feature for writers because they need to read them, to benefit from them. In reverse outlining, this in itself is considered useful: \citet{tully_reverse_2019} noted that seeing reverse outlines gives a ``fresh eye'' on the text as it motivates taking writing ``breaks'' and ``time away from a draft weakens the memory's tie to the narrative'' -- thus stimulating (self-)reflection. Since writers in our concept do not write summaries themselves, it may be even easier for them to achieve this ``fresh'' view through the summaries. Indeed, we found this in our study (\cref{sec:results_reflection_strategies}).

\subsection{UI: Summaries as Margin Annotations}

The writing interface mimics known writing software and is split into two main views. One view represents a classic word processing editor. The other view represents the margin, the usual place for annotations. To stay in line with common terminology in user interface design, we termed this second view the ``sidebar''. 
Writing guides suggest using the document margins or separate notes (or post-its) for additional text annotations, for example to develop a reverse outline (e.g.~\cite{lab_reverse_2021, uni_tennessee_writing_strategies_2007}), or as ``sideline commentary''~\cite{levan2017self}. Concretely, we represent each paragraph as a card, based on the suggestions of \citet{messuri_revision_2016} to cut summaries on paper into multiple snippets. %
The details of this from an interaction point of view are covered below (\cref{sec:interaction_concept_cards}).
The text editor and the sidebar are inherently linked by their content. Each text paragraph in the editor produces a summarised annotation card in the sidebar. To support this, we highlight the paragraph when clicking on its card (cf.~\cref{fig:teaser}). %

\subsection{Interaction: Summaries as Cards}\label{sec:interaction_concept_cards}

Within the sidebar, the summaries are presented as ``cards''. This is motivated as a metaphor considering the use of pen and paper in (reverse) outlining and other revision activities (e.g. margin notes, post-it stickers). There, notes may be written on cards (or post-it notes) to afford a set of relevant interactions, which we map to interactions with our digital cards as described next. In each part, we indicate the mapping of \textit{physical (paper) action $\rightarrow$ UI interaction}.

\subsubsection{Reordering: Move paper notes $\rightarrow$ Drag \& drop cards}
Related work has commented on the benefit of reordering paragraphs in the writing process \cite{duke_university_revising_2021, messuri_revision_2016, messuri_writing_2016, saltz_harvard_1998}. The underlying strategic motivation and intention is that the ability to step back from the text to reorder paragraphs supports and enhances structural revision. %
With a (reverse) outline on a physical piece of paper this means cutting the outline paper into smaller snippets (if not already written on cards) so that these can be moved around and arranged in a new order. In our UI, we mapped this to cards (i.e. summaries are ``precut'' per paragraph) and the interaction of dragging and dropping these cards vertically in the sidebar. 

\subsubsection{Removing: Throw paper away $\rightarrow$ Delete card}
Removing content is one possible action that writers might identify when reflecting on a (reverse) outline of their text~\cite{duke_university_revising_2021, hamburger_outlining_2013, saltz_harvard_1998, uni_tennessee_writing_strategies_2007}. For example, this might be triggered by the insight that some parts of the text are redundant. With paper notes, writers can remove said note while in our UI we provide a ``delete'' button for each card (see \cref{fig:teaser}).

\subsubsection{Splitting: Cut/rewrite paper notes $\rightarrow$ Add line break}
Another related action is splitting, which writers might identify as a way of improving their text based on insights they gained from a (reverse) outline~\cite{duke_university_revising_2021, messuri_revision_2016, messuri_writing_2016, university_of_wisconsin-madison_reverse_2021}. For example, they might see that one paragraph mixes two topics. On paper, writers may cut a note (or rewrite it on two new pieces of paper). In our UI, writers can simply add a new line break in the main text to split a paragraph into two, which automatically updates the cards in the sidebar accordingly.

\subsubsection{Merging: Glue/rewrite paper notes $\rightarrow$ Delete line break or drag \& drop cards onto each other}
Inversely to to splitting, (reverse) outlines may also reveal to writers that two text parts should be merged into one~\cite{university_of_wisconsin-madison_reverse_2021}. On paper, writers might glue notes together or rewrite them onto a single new card. We support this in two ways: For a simple concatenation, users can move the involved paragraphs next to each other (if they are not already subsequent) and delete the line break between them, which also updates the cards accordingly. Alternatively, if the merge is semantically more involved, users can drag \& drop one card \textit{onto} another one, which will trigger a dedicated merge view with an automatically created merge suggestion (see \cref{fig:merge_view} and \cref{sec:implementation} for details). Users can accept or cancel this suggestion. If they accept it they can of course also further edit the result in the main text view.

\subsubsection{Revision: Rewrite/reprint main text $\rightarrow$ Copy card content}
Finally, writers might sometimes identify content in their annotations as suitable pieces for a revision of the main text, for example, to achieve a more succinct version. On paper, writers could rewrite/reprint the main document or manually cross out or write over it. In our UI, we provide a button to replace a paragraph directly with its summary (in version 1 of our prototype), which we later turned into the more flexible concept of copying the content of a card to the clipboard (in prototype version 2). We report more details on this conceptual change in our results and discussion.

\subsection{Annotation Content and Controls}%
\begin{figure*}
    \centering
    \includegraphics[width=0.8\textwidth]{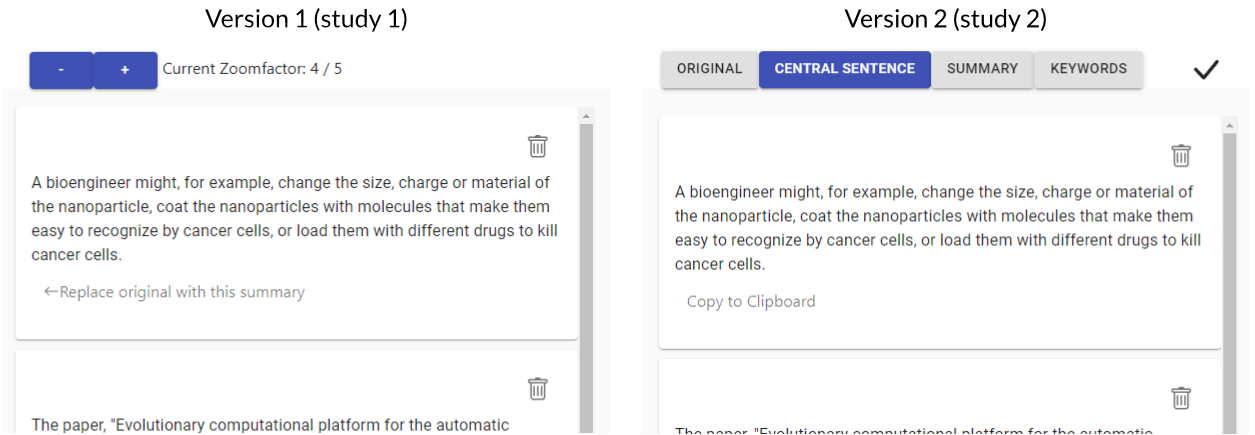}
    \caption{Our design iteration of the summary controls, from version 1 (used in study 1) to version 2 (used in study 2): We redesigned the summary levels (at the top of the UI) and replaced the ``Replace in Text'' action with a more flexible ``Copy to Clipboard'' action (at the bottom of each card). We also added an indicator in the top right that shows whether summaries are currently being updated in the background (spinning circle or checkmark).}
    \label{fig:comparison_of_ui_versions}
    \Description{A screenshot showing two iterations of the sidebar view. The first iteration shows a -/+ button to change the zoom factor. The final iteration shows four separate buttons with concrete labels for the zoom factors: original, central sentence, summary, keywords.}
\end{figure*}

Another key aspect of our concept concerns the content of the annotations/cards. Many choices are possible here, considering different kinds of summaries (\cref{sec:related_work_summaries}) and degrees of granularity. We thus offer the user control over multiple levels. Providing more than a single option in this way is further motivated by varying comments on length and type of summaries for reverse outlining in writing guides (e.g. one sentence, main idea) \cite{tully_reverse_2019, hamburger_outlining_2013, lab_reverse_2021}. In our protoype, we explored two sets of levels: 1) Abstract numbers ranging from one to five representing \textit{increasing summary zoom levels} and 2) descriptive summary levels (see \cref{fig:comparison_of_ui_versions}). %

\begin{table}[]
\centering
\footnotesize
\newcolumntype{L}{>{\raggedright\arraybackslash}X}
\renewcommand{\arraystretch}{2}
\begin{tabularx}{\linewidth}{LLL}
\multicolumn{1}{c}{\textbf{UI Element}} & \multicolumn{1}{c}{\textbf{Version 1 (study 1)}} & \multicolumn{1}{c}{\textbf{Version 2 (study 2)}}\\
\midrule
Cards-text link & Click on card navigates to paragraph & Added text highlighting when card is clicked (see~\cref{fig:teaser}) \\
Copy summary (compare in~\cref{fig:comparison_of_ui_versions}) & ``Replace original text with this summary'' & ''Copy to clipboard'' \\
Merge & No visual highlight of merged content & Added visual highlights of retained and cut parts of the merged paragraphs (see~\cref{fig:merge_view}) \\
Summary levels (compare in~\cref{fig:comparison_of_ui_versions}) & \textit{Level 0:} original paragraph in full; \textit{Levels 1 - 4:} extractive summary with (4, 3, 2, 1) sentence(s); \textit{Level 5:} abstractive summary & \textit{Original:} full text; \textit{Central sentences:} extractive with 1 sentence; \textit{Summary:} abstractive; \textit{Keywords:} extract up to 5 keywords\\
\bottomrule
\end{tabularx}
\caption{Prototype changes implemented in Version 2 after the feedback from study 1 (Version 1).}
\label{tab:comparison}
\end{table}

\section{Implementation}\label{sec:implementation}
Our prototype has a server client architecture, with the front end UI, a Python server to host the text summarization methods, and a webserver and database to serve the web app and questionnaires and store data collected in the studies. %
We improved the prototype after the first study (described in detail in \cref{sec:results_design}). \cref{tab:comparison} summarizes the differences between the two versions.

\begin{figure*}
    \centering
    \includegraphics[width=0.7\textwidth]{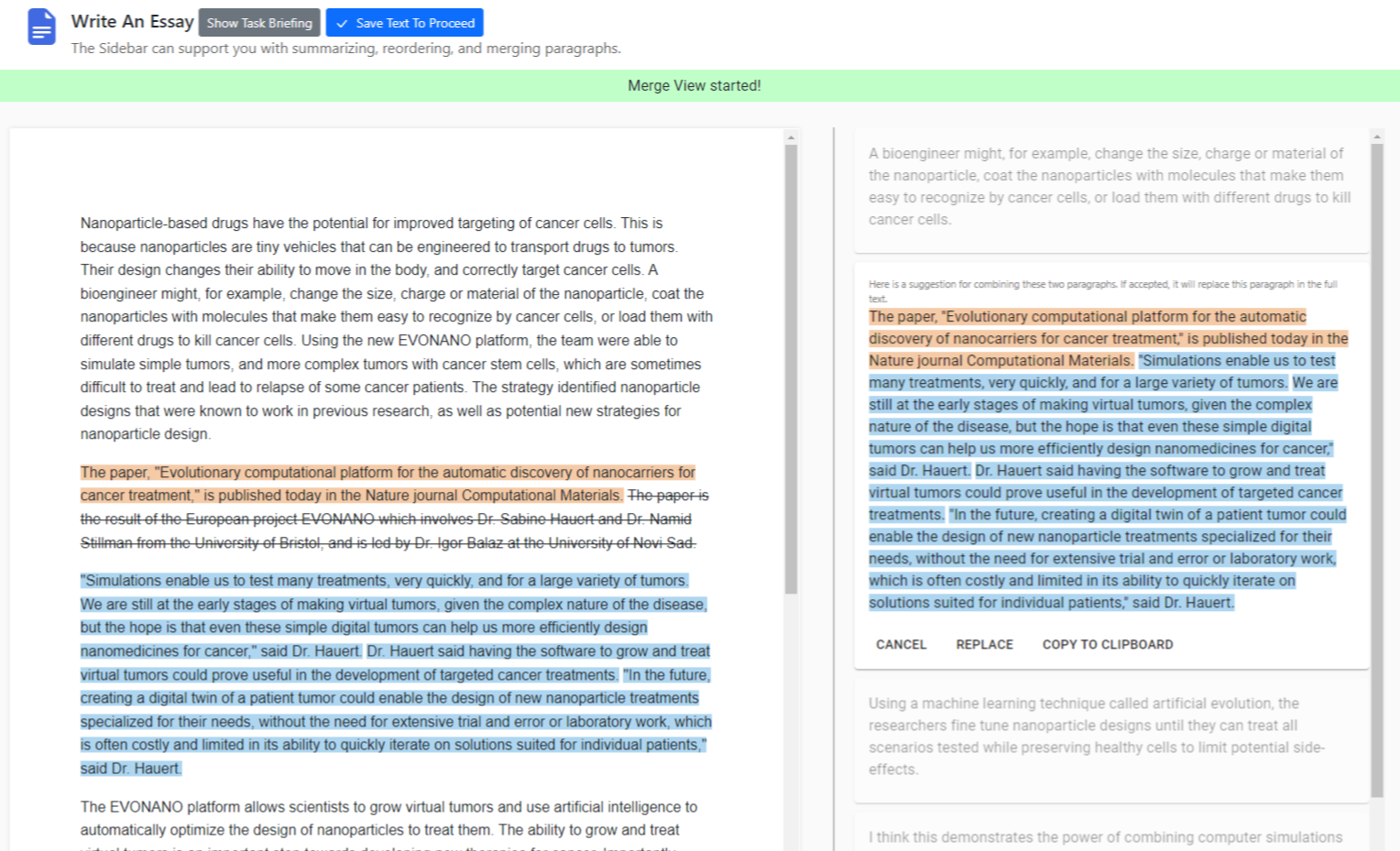}
    \caption{Merge view (in prototype v2 / study 2), triggered by dragging one card onto another. The sidebar shows a merge suggestion in a merged card. A green message (top), removed summary controls and grayed out other cards indicate that no further interaction is possible until the merge is accepted or canceled. The suggestion is computed by ranking sentences in each paragraph (\protect\cref{sec:implementation}). The page view highlights retained and cut parts of the merged paragraphs (color-coded, plus strikethrough for cuts). Version 1 was the same but showed no colors and strikethrough.}
    \label{fig:merge_view}
    \Description{A screenshot showing two highlighted text paragraphs. It shows the state after two cards in the side panel have been merged. The highlighting in the main editor and side panel show the same color.}
\end{figure*}

\subsection{Frontend / UI}
We implemented the UI (\cref{fig:teaser}) as a web app with React.js. It realizes the concept described in \cref{sec:concept}: It is split into a main text editor and a sidebar with summary cards. Each paragraph is represented by one card. The cards can be reordered, deleted or merged.

\subsubsection{Card Interactions}
We further implemented all card interactions as described in \cref{sec:concept}. Card dragging was implemented using react-beautiful-dnd. %
In version 2, clicking a card highlighted its paragraph in the text in color for one second (green box in \cref{fig:teaser}). %

\subsubsection{Merge}
For the merge functionality, the user drags and drops a card onto another card. The system then applies extractive summarization to rank the sentences in each paragraph. The merge suggestion is composed by concatenating the five top ranked sentences across both paragraphs. The original order of these sentences in the text is preserved. In version 2, the page view highlighted retained and cut parts (see~\cref{fig:merge_view}). %
A green message, removed summary controls and grayed out cards indicate that no further interaction is possible until the user has clicked ``cancel'', ''replace'' (which applies the merge in the main text directly) or ''copy to clipboard''.

\subsubsection{Summary Levels}
The toolbar at the top provides summary options, which we redesigned after study 1 (\cref{fig:teaser} shows the improved version; for a comparison see \cref{fig:comparison_of_ui_versions}). \cref{tab:comparison} describes the options in each version. %
\cref{sec:implementation_nlp} describes the NLP methods.

\subsection{Language Models and NLP Methods}\label{sec:implementation_nlp}
Here we describe the used language models and NLP methods. We made pragmatic model choices for our interactive system since our focus was the writing insights. Concretely, we chose a tradeoff between quality and computation speed, as assessed by trying out related models of different sizes. 

\subsubsection{Extractive Summarization}
The extractive method uses \textit{GloVe} embeddings\footnote{\url{  https://nlp.stanford.edu/projects/glove}} and \textit{TextRank}~\cite{Mihalcea2004textrank} to find the top k sentences. In study 1, the extractive method was used for summarisation Levels 1 to 4, beginning with k=4 sentences on Level 1 with k decreasing by 1 per level until k=1 on Level 4. In study 2, extractive summarisation was used for the Central Sentence level with k=1.

\subsubsection{Abstractive Summarization}
Our abstractive method used the \textit{T5} transformer model~\cite{Raffel2020} available via \textit{huggingface}\footnote{\url{https://huggingface.co/t5-base} -- These T5 transformer models are trained for summarization and T5-base refers to a medium model size.}, with the language modeling head.\changenote{We mostly relied on defaults but determined these ``generate'' parameters by early explorations with our prototype: \textit{num\_beams}=4 , \textit{no\_repeat\_ngram\_size}=2, \textit{early\_stopping} was enabled and \textit{max\_length} was set to 70\% of the source token count.} In study 1, abstractive summarization was used in Level 5. In study 2, the level named \textit{Summary} used it.

\subsubsection{Keyword Extraction}
We implemented the keyword extraction using the open source library \textit{wordwise}\footnote{\url{https://github.com/jaketae/wordwise}} (which uses a RoBERTa model\footnote{\url{https://huggingface.co/docs/transformers/model_doc/roberta}}).
Keyword extraction was not implemented in study~1 but was used for the \textit{Keyword} level in study~2.

\subsection{Backend / Server}
The prototype was hosted on a university server with 32GB RAM and a GPU with 12GB memory. %
We implemented a lightweight Python Flask API to process the text. Each method's backend call receives a JSON Array containing all of the required paragraphs and returns the desired results as a JSON object with each index representing one paragraph. We created a cache in order to improve loading and processing times. It stores the paragraphs as well as the return values of each summary technique that was previously used on that paragraph. It also records which paragraphs have been modified as a result of the author's revisions to the original text. Thus, the system only needs to newly compute the summaries of changed paragraphs. Overall, the resulting summary updates were almost instantaneous for extractive methods and keywords, and took up to 2 seconds (when adding long text at once i.e. via copy/paste) for the abstractive method.

\section{Method}

Our evaluation methods consist of a writing task with think-aloud, followed by a semi-structured interview, all during online sessions (video calls), plus a final questionnaire. For better clarity, we describe these methods upfront here. The study procedure involving these methods is then presented in detail in \cref{sec:study}. %

\subsection{Think-Aloud During Interaction}
Participants interacted with the prototype in a writing task, for which we asked them to articulate their thoughts. Occasionally, we also asked questions to better understand their thinking (e.g. when they seemed to be looking for something) or to remind them of thinking aloud. %
We recorded these video calls including audio and participants' screens. On top of that, two researchers involved in the video call took unstructured notes on comments from the participants as well as critical observations. Both researchers later shared, compared and merged their notes.

\subsection{Semi-Structured Interviews}\label{sec:methods_interview}
A semi-structured interview followed the writing task to dig deeper and allow people to share thoughts in reflection on their experiences. We asked questions about notable moments we observed during the task, plus detailed questions about which aspects of the prototype people particularly liked or disliked or would prefer to change.

\subsection{Coding of Think-Aloud \& Interviews}\label{sec:methods_think_aloud}
We used the researcher notes from the online sessions and transcripts of people's comments in an approach adopting Grounded Theory~\cite{corbin1990basics, muller_grounded_2012}: In an \textit{open coding} round, two researchers individually assigned inductive codes to each note. %
In an \textit{axial coding} round, these two researchers compared their codes and clustered them into higher thematic groups. For example, emerging clusters grouped several codes that related to \textit{improvement of the prototype} or codes that related to the \textit{interaction strategy} of the participants. Afterwards, the researchers jointly agreed on a ``cluster label'' that best describes its corresponding codes. These resulting aspects also serve us as a structure for the results in this paper (\cref{sec:results}). Finally, in a \textit{selective coding} round, three researchers went through the full transcripts of all interviews to find further evidence (and potentially counterexamples) for these aspects as developed in the previous step. Since we had created the transcripts automatically, we double-checked all relevant parts again in the original videos.

\subsection{Questionnaire}\label{sec:methods_questionnaire}
In an online questionnaire after the video call, people rated each feature on a five-point Likert scale (plus a \textit{did not use} option). 
Two open questions asked about (1) other writing use cases where a system such as ours might be useful, and (2) what participants would like to add or change in the current prototype. %

\section{User Studies}\label{sec:study}

We conducted two user studies via video calls, with a design iteration of our prototype in between them. The procedure was identical in both studies. To avoid redundancy, we report on it here once.

\subsection{Participants}
In total, N=12 people (5 female, 7 male) participated in the two studies (4 in the first study, 8 in the second one; no one participated in both studies). Their age ranged from 22-36 years. 
\changenote{
We recruited them from networks across a few universities and personal contacts. Their backgrounds included professionals, undergraduate students, and (HCI) researchers to whom we reached out via email.}
While this is a convenience sample and we reflect on limitations of the study in our discussion, our sample covers relevant users for writing tools with  an interesting range of (professional) writing experiences and regularity. Concretely, about half of participants indicated to write (substantially) daily, the others less than once a week.
Participants were not native English speakers yet all used English regularly professionally and/or personally and were informed beforehand that they would be asked to write an essay in English. We further informally confirmed their high proficiency based on the task observations and resulting texts. %
People were compensated with a 10 Euro gift card for an online shop.

\begin{figure}
    \centering
    \includegraphics[width=\linewidth]{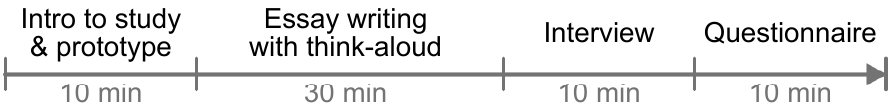}
    \caption{Overview of our study procedure (see Section~\protect\ref{sec:study_procedure}).}
    \label{fig:study_overview}
    \Description{A graph showing the four phases of the user study: Introduction, essay writing, interview, questionnaire.}
\end{figure}

\subsection{Procedure}\label{sec:study_procedure}
On average, a study session took 60 minutes. It was structured as shown in \cref{fig:study_overview}. We explain the involved steps in more detail next. %

\subsubsection{Study Intro (10 minutes)}
In line with our institutional regulations and informed consent procedures, an intro page in our web app explained the study, informed about data collection and privacy regulations, and further general study information. One of the researchers then demonstrated the features of the prototype as an introduction via screen sharing. The text used in this demonstration was taken from an arbitrary daily article on Wikipedia. As part of this demonstration, the researcher showed the different summarization levels, the merge functionality, and the linking between cards in the sidebar and paragraphs in the page view. People could ask questions, for example to clarify how the interactions worked.
    
\subsubsection{Essay Writing (30 minutes)}\label{sec:procedure_writing_task}
In the main part, people chose one of two opinionated articles and wrote an analytical essay on how the respective author builds their argument. The task prompt and articles were taken from the analytical writing section of practice test prompts for the Standardized American Tests (SATs)\footnote{https://blog.prepscholar.com/sat-essay-prompts-the-complete-list (Topic 1: Benefits of early exposure to technology, Topic 2: Preservation of natural darkness)}. We chose this task because it allows users to work with an existing text and thereby increases the opportunity to experience the prototype without an extensive ``creative'' drafting period. At the same time, this task requires people to analyze and write about the text, not to simply copy it, thus ultimately leading them to write an essay. We deemed this a useful trade-off between starting from scratch and starting with a given text in the prototype (i.e. pure editing task). We reflect on this choice in our discussion. People opened the article in a separate browser tab so that they could switch between the prototype and the article. %
We encouraged thinking aloud (see \cref{sec:methods_think_aloud}). With people's consent, we recorded audio and screen during the essay writing and the following interview. 
    
\subsubsection{Concluding Interview and Questionnaire (10+10 minutes)}
We conducted the semi-structured interview after the writing task (see \cref{sec:methods_interview}).
Finally, after the video call, people filled in the questionnaire (see \cref{sec:methods_questionnaire}). %

\section{Results}\label{sec:results}

\begin{figure}
    \centering
    \includegraphics[width=\linewidth]{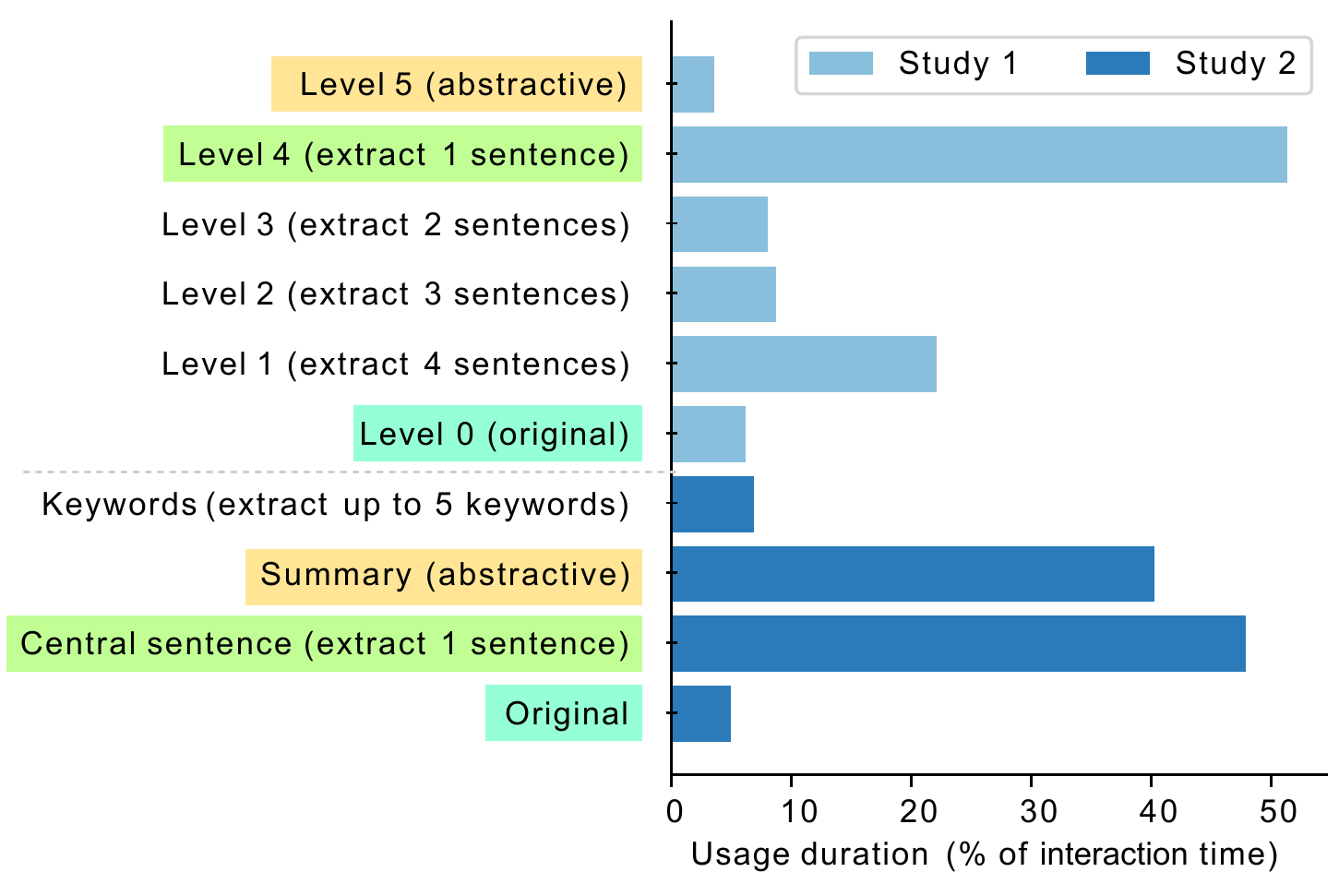}
    \caption{Time spent in each summary setting. Colored labels show which settings from prototype version 1 (study 1) correspond to which settings in version 2 (study 2). Uncolored settings only appeared in one design/study.}
    \label{fig:usage_durations}
    \Description{A graph showing how much time participants in study have spent in each summary setting. The graph is split into the first iteration and the final iteration.}
\end{figure}

We structure this report into design insights and identified interaction, writing and reflection strategies.
The mean duration of interaction and interviews was 40 minutes, matching our planned time for these parts. The mean final text length was 416 words. All summary settings were used and both extractive and abstractive methods were used considerably in the revised design (see \cref{fig:usage_durations}).

\subsection{Design Insights and Improvements}\label{sec:results_design}%
We report findings in study 1, that motivated prototype changes before study 2. We report these together with a description of the changes and observed consequences in study 2. %

\subsubsection{Unclear Semantic Zoom and Summarization Levels (Study 1) were Redesigned (Study 2)} 
The initial zoom metaphor was a source of confusion. %
First, the minus/plus symbols of the zoom tool where interpreted as text length, which is the inverse mental model: As text length, ``minus'' should decrease summary length but it was designed as ``zooming out'' semantically (i.e. seeing more of the original text, thus actually increasing length). 

Second, people had trouble understanding how the levels differ: \pquote{I felt level one to three did not show much difference [...], this may also be because I included little information}{1}{1}\footnote{Participants' quotes were translated into English by the authors.}.  %
Indeed, for short paragraphs, the first three zoom levels may not result in text changes because extracting a varying number of sentences does not make a difference if a paragraph has fewer ones anyway: %
\pquote{I didn’t realize that there were two more levels [...] since nothing changed, I felt that the few paragraphs I have created may not have been sufficient to trigger something here}{4}{1}. %

These issues motivated us to redesign this tool: With version 1, we explored a rather fine-grained set of summarization options. However, as described above, this was not particularly helpful for people. We thus decided to only keep four levels in version 2, also motivated by what was used the most (\cref{fig:usage_durations}): \textit{original} text paragraph, single \textit{central sentence}, short \textit{abstractive summery}, and extracted \textit{keywords}. We also labeled these levels with text captions on four separate buttons, which avoids the plus/minus. These changes were clearly successful: We found no indication of misunderstandings in both observed usage behavior and think-aloud comments. %

\subsubsection{Unclear Correspondence of Text and Cards (Study 1) and Improvements (Study 2)}\label{sec:results_design_linking}
Participants in the first study often commented on ``losing orientation'' when interacting with the cards: For example, some people had issues deciding which card matched which text paragraph at a glance.

We partially solved the issue by highlighted the corresponding text paragraph when a user clicked a card. All participants in study 2 appreciated this; for example: \pquote{It’s a good feature to be able to highlight this, then it is easier to find [corresponding text]}{11}{2}. %
However, this highlighting on click was only from cards to text. As some people's behavior in study 2 showed, it should be bidirectional (i.e. clicking on paragraph to highlight its card). For example, as a workaround, one person (\participant{6}{2}) quickly clicked through the cards until their paragraph of interest was highlighted. %

A related issue in version 1 was the button in each card to replace content in the text (see version 1 in~\cref{fig:comparison_of_ui_versions}, bottom of card). %
The resulting change lacked visual feedback: \pquote{I thought that something was happening [...] but suddenly it’s gone [the text paragraph] and something has changed}{3}{1}. %
Considering further observations, we solved this problem not with more feedback here but more generally by changing this feature to \textit{Copy to Clipboard} (see version 2 in~\cref{fig:comparison_of_ui_versions}), as described next (\cref{sec:results_copy_to_clipboard}).

\subsubsection{Unflexible Direct Changes (Study 1) were Replaced with a Copy to Clipboard Feature (Study 2)}\label{sec:results_copy_to_clipboard}

In the first version and study, the \textit{Replace in Text} button in each card replaced the corresponding paragraph with the card's summary directly. This had three downsides: (1) The replacement was inflexibly tied to the summarized paragraph -- it was not easily possible to use the summary elsewhere in the text. (2) As mentioned above, the lack of visual feedback was confusing. (3) Finally, this button encouraged people to view the summaries as text \textit{suggestions} rather than as annotations. 

We thus change this to a \textit{Copy to Clipboard} button. This requires users to perform an extra step (paste) to include the summary into their writing. This offers more freedom to decide where in their text they want to insert it. At the same time, as we indeed confirmed in study 2, this change discouraged people to view the summaries as text suggestions that are presented to be ``accepted''.

\subsubsection{Difficulty of Assessing Merge Changes (Study 1) was Addressed with Better Merge Preview (Study 2)}\label{sec:results_merge}
People in study 1 initially found it difficult to spot the differences of the summarized text and the original paragraph. Especially when merging two paragraphs they wanted to see more clearly which parts of the text were kept in the merge suggestion and which were removed. The second iteration of our prototype thus highlighted this with colors and font styling (see \cref{fig:merge_view}). %
This was a successful change as people in study 2 commented positively on this visual feedback. %
    
\subsection{Interaction and Writing Strategies}\label{sec:results_writing_strategies}

Here we report on key interaction patterns and writing strategies. %

\subsubsection{Using Automated Summaries to Understand Longer Text}\label{sec:results_summaries_as_overview} 

One distinct strategy was to gain a quick overview of longer text, for example, by copying parts of the source article into our editor and reading the summaries: %
\pquote{Let’s see what the editor says about the individual paragraphs. [I’m going to] look at the text summaries, or more concretely, the central sentences in order to get a better overview}{3}{2}. %
This was perceived as fast: %
\pquote{It is definitely faster because one does not have to reread everything}{8}{2}.%

Related, summaries simplified the text for reading and selection for building an argument: %
\pquote{I wanted to look at a shortened version, to see whether the summary [of a part of the source article] is irrelevant for my argument}{6}{2}. %
As can be expected, this strategy was employed at the beginning, prior to writing. It started with trying out summary levels before settling for one for reading. The strategy ended after reading the summaries and was followed by two distinct transitions to writing: (1) One group integrated the summaries directly into their text as a starting point. (2) Others deleted the article form the editor again to start with a blank page. We describe the larger strategies emerging from this next (\cref{sec:results_topdown_bottomup}).

\subsubsection{Developing Text Top-down vs. Bottom-up}\label{sec:results_topdown_bottomup}
Two larger distinct writing strategies emerged in our study: 
First, in what we call \textit{bottom-up} text development, participants started with an empty editor and then built up their text through typing, interleaved with checking the reference article for the task. Occasionally, they copied specific snippets from this source article into the editor to read the summaries and/or use the text as a basis for further writing. 

In contrast, in what we call \textit{top-down} text development, participants copied the entire source article into the editor and used the summaries as text building blocks. In this strategy, people kept the source's argumentative structure and mainly focused on drafting sentences to connect the summary-based paragraphs.

Across both these strategies, people edited summaries (or directly used source text) to avoid plagiarism and to match their own writing style. We describe this behavior in more detail next (\cref{sec:results_leaps}).

\subsubsection{Adapting Text as Part of Integrative Leaps and ``Reverse Leaps''}\label{sec:results_leaps} 

When using text from summaries to revise or draft text, the most dominant strategy was modifying the summary text to match one's own writing: %
\pquote{This is not copy-pasted one-to-one. I have partly adapted it [the copied summary] and reformulated it}{6}{2}. %
These edits can be seen as \textit{integrative leaps}, as recently introduced by \citet{elephant_tochi2022}: It describes people's efforts to integrate (AI) suggested material into their writing. 
We also made the \textit{opposite} observation where people adapted their own writing to influence the summaries: For example, one person combined multiple text paragraphs to explore changes in the summaries. Similarly, one person said: \pquote{I'll remove all empty spaces between the paragraphs to see whether the message I want to deliver comes through}{3}{1}. This also demonstrates reflection on paragraph structure and scope, which we elaborate more on in~\cref{sec:results_check_paras}.

\subsubsection{Summaries as Textual Building Blocks for Overview Texts}
People further considered the text from automated summaries to write ``overview sections''. 
For instance, one person framed this as building blocks to formulate their text's conclusion: \pquote{I want to add final summary sentences at the end, [...] like a conclusion. I quickly skimmed through existing text summaries to recall what the main messages were}{5}{2}. While another 
participant mentioned a similar strategy to build an abstract: \pquote{I'd take the entire text and try to generate an abstract out of it}{11}{2}.

\subsection{Reflection Strategies}\label{sec:results_reflection_strategies}

Here we ``zoom in'' on people's reflective use of our system. %
        
\subsubsection{Considering Summaries as Another, External Perspective}\label{sec:results_reflection_external_perspective}

A key aspect that characterizes the reflection processes with the summaries is that they were considered as an external view on the writing; such as: \pquote{And that’s how I worked with [the summary], I’d first write down my text and would then compare: What does the bot suggest? And then I’ve integrated [the summary] in my text}{11}{2}. And: \pquote{It’s good to read [the paragraph], in other wording, and [...] slightly summarized}{12}{2}. %

Related, also hinted at in the first quote above, people \textit{compared} the AI summary with their own ``mental summary'' for the same paragraph. Noticing differences prompted people to more closely analyze why this was the case. In doing so, they reflected on their written work. For example: 
\pquote{[The key message] is essentially what the tool also proposes as a summary. [...] It's interesting that what the tool proposes is longer than my own summary}{9}{1}.
This reflection often also led to edits. %

\subsubsection{Using Summaries to Check and Revise Paragraph Scope and Structure}\label{sec:results_check_paras}

Another way of using the summaries for reflection was to check the scope of individual paragraphs, for example to determine if one indeed included its intended content or message. 
For example: \pquote{I would take a look at both [the paragraph and its summary]; both things are important. I would like to see in the summary whether all my arguments are also reflected [...]. I’d expect the summary to help me reflect where I set my focus in the section [...]}{5}{2}. %
If the summary did not include the intended aspect, people applied two revision strategies: (1) Either the main message is elaborated on in the paragraph, or (2) the paragraph is split.

Related, restructuring actions were also valued: %
\pquote{The most helpful [features] were definitely the reorder and merging of paragraphs. When I noticed that the summaries at a specific zoom level were too short, I could just merge them; that simplified the work and made work much faster. That was the most practical aspect. Other than that, the summary and simplification of large paragraphs}{1}{1}. %

\subsection{Feedback in the Final Questionnaire}

The final questionnaire asked people to think about where (else) they would use this system. The answers echoed the topics of overview and reflection identified during the interaction and interviews: \pquote{[...] it could be helpful for organizing my chapters and for getting a quick overview of different parts}{2}{1}. And another person thought it would be useful \pquote{[w]hen writing texts where structure and conciseness matters (blog entries, news articles, summary/conclusion [...])}{5}{2}.

The questionnaire also asked about feature ideas and changes. We addressed the replies from study~1 in our design iteration (e.g. highlighting text changes, cf.~\cref{sec:results_merge}). Replies from study~2 revolved mainly around further improving the visual link of paragraphs and summaries (e.g. one person suggested to highlight the paragraph already on hover, not click).

\begin{figure}
    \centering
    \includegraphics[width=\linewidth]{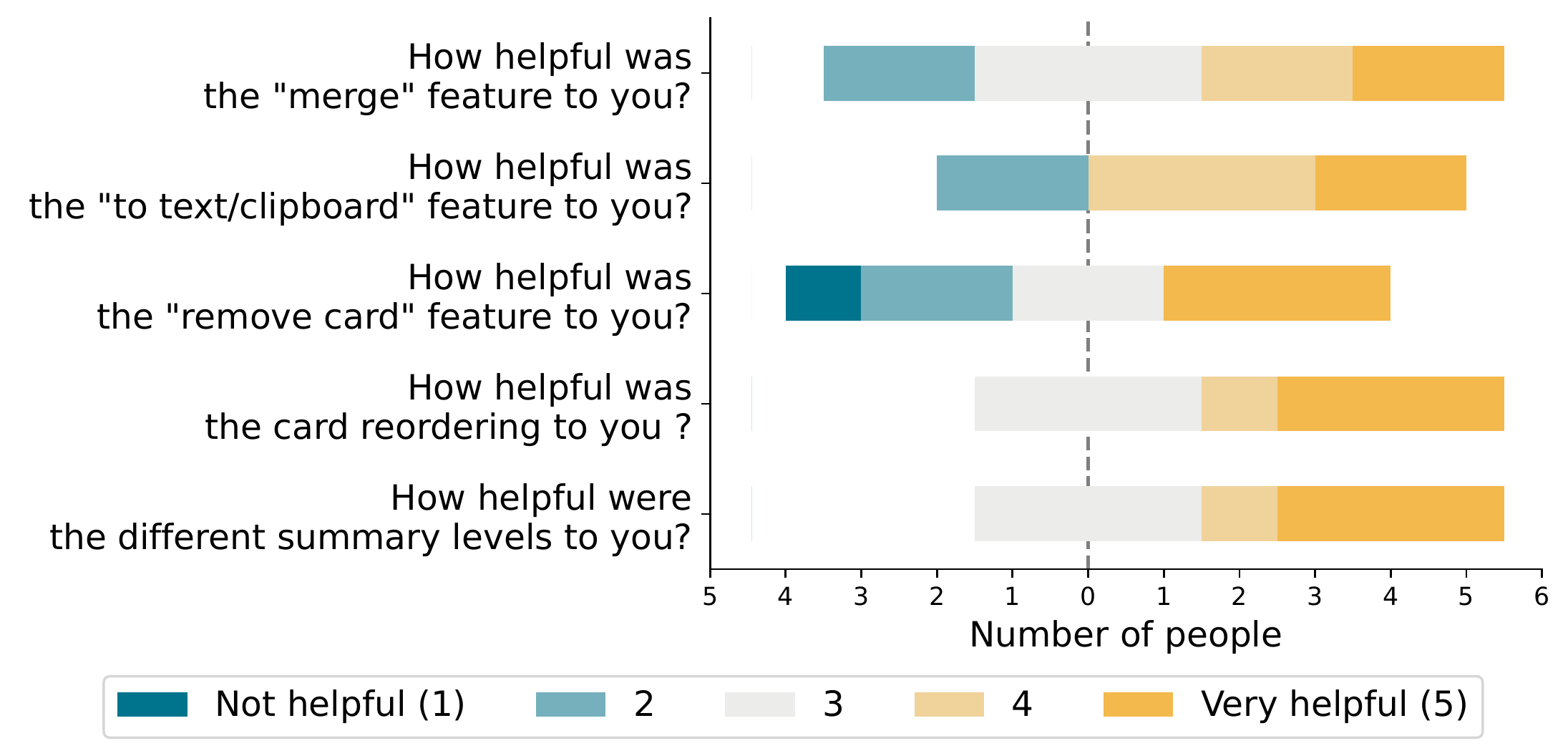}
    \caption{Likert ratings of specific features asked in the final questionnaire. The number of ratings varies because a ``did not use'' option was available for each question.}
    \label{fig:likert_results}
    \Description{A graph showing participant's responses to the Likert questions.}
\end{figure}

\cref{fig:likert_results} shows the results of the included Likert items on individual features. Overall, not everyone found every feature equally helpful (or used it) yet all features were helpful for a considerable proportion of people. This fits to the observed diversity of people's writing approaches. ``Delete'' was the least helpful feature.

\subsection{Comparisons against Writing without Automatic Summary}\label{sec:results_comparison}
\changenote{
 Even without a comparative baseline a few participants explicitly commented on their writing experience with and without automatic summaries.
}
\changenote{
Their comments hint at the use of different kinds of manual summarization in their usual work flows, and how these benefit from our tool. 
For example, a common reading technique involves highlighting important parts of the text, which corresponds to our ``central sentence'' level:
\pquote{I would have just started writing [...] and summarized the relevant paragraphs from the source in the process. I would have highlighted important parts in the text and only after finishing the entire text I would come back to think about the structure of the text.}{8}{2}.
}

\changenote{
In another example, a participant explained that %
\pquote{[w]ithout the margin summaries, I would probably first draft the entire text finally to recognize that my submission was too long [...] only after [that] I would start summarizing and shortening the text}{4}{1}.
}

\changenote{
Finally, automatic summaries shorten reading time when writing with source material, as in the study: %
\pquote{[...] it really formulates concretely what I need. In the end it is exactly what I would have usually done laboriously by hand. I would read for couple hours. Here I can just copy the text.}{9}{1}.
}

\section{Discussion}\label{sec:discussion}

\subsection{How Continuous Automatic Text Summaries Support Writers}

Writing is an iterative process interleaving several sub-processes that relate to planning, drafting and evaluating text~\cite{flower1981cognitive}. Our continuous automatic text summaries consider this in that they are updated as users write their text and thereby succinctly reflect the state of the draft, as an offer for evaluative processes. This concept was accepted by participants and utilized in diverse ways: Crucially, the summary annotations were used and valued by people as a support for reflection and overview. Concretely, people reflected on the content, scope and structure of their paragraphs, by integrating insights from the corresponding summaries. This also matches the intention of the reverse outlining strategy, which we had used as a starting point for our design. %
Moreover, the summaries helped writers to gain a quick overview of both the given text at the start as well as their own essays at the end. Finally, as a key finding on the perception and human-AI relationship here, people viewed these summaries as another, external view on the text. It is worth noting that this perception was not limited to summaries of text copied from the given articles but also was the case for summaries of the users' own writing. This ``external AI view'' can be helpful, especially when otherwise writing alone. %

\subsection{Self-Annotation and ``AI Annotation''}

We critically reflect on the shift from self-annotation to ``AI annotation'' in this work.
First, manual summarizing can be helpful as part of (meta)cognitive strategies for writing, text understanding and learning to write~\cite{waters_metacognition_2009}. Automatic summaries might take this away. However, participants indeed found the summaries helpful, as described above.
Interestingly, a typical behavior involved \textit{comparing} a mental summary to the generated summary. This indicates that AI annotation influences cognitive processes in self-annotation, instead of replacing them. Crucially, as pointed out previously, AI annotations were to some extent considered as an external view. Future studies could further relate this to perceived agency and authorship, as examined for writing with text generators (e.g.~\cite{Buschek2021chi, yang_ai_2022}). Notably, low perceived authorship \textit{for annotations} might have positive utility for writers (i.e. as ``fresh eyes''~\cite{tully_reverse_2019}, as suggested in our study) -- in contrast to the case when generating or editing text, where AI might be seen as taking away agency. %

A second point is expressiveness: Self-annotation offers more flexibility than our UI, both spatially and symbolically, (e.g. freehand drawings, also beyond the margins). The spatial aspect appeared in our study in comments on locating annotations in the main text (cf.~\cref{sec:results_design_linking}) yet people did not seem to miss or expect visual annotations as part of our concept. Our takeaway here is that the \textit{correspondence} of AI annotations and user content needs to be clarified explicitly (e.g. our click-to-highlight design change) because people lack the ``cognitive context'' that would come implicitly via manual annotation. At the same time, users' expected \textit{modality} of NLP-powered annotations in our design was clearly textual. %

Combining these two points, our findings guide the community towards exploring AI annotations as \textit{complementing} self-annotation with two concrete directions: (1) supporting a mix of manual and automated annotations (e.g. reusing UI concepts from human-human collaboration, such as differently colored annotations for writer and AI), and (2) annotations beyond text (e.g. by combining our work with the stylus diagramming interactions by \citet{Subramonyam2020} or even text-based sketch generation systems~\cite{Huang2020}).

\subsection{Limitations \& Reflections on Methodology}

We do not claim that the generated summaries represent the latest state-of-the art in NLP. Although people sometimes had reservations about summarization quality (e.g. fearing at the start that it might lack important context), they productively worked with the summaries and also integrated them into their own text. \changenote{From our observations we hypothesize that even with ``perfect'' quality users would experience AI summaries as external because they are not written by users themselves, need to be read, and trigger comparison to own expectations. Related, in future work, it might be interesting to examine the impact of how well AI summaries match a user's writing style.}

It is challenging to let people experience not only free writing but also revision within the limited duration of an observed study session. Here, we reflect on lessons learned:
First, our observations and people's feedback on the study suggest that writing experience may influence how comfortable people feel with being observed. We believe our task was useful here because it gave people something concrete to get started with, while familiarizing themselves with the study situation, without having to write immediately (i.e. reading the provided article first). %

Second, however, the task of writing about a single article does not perfectly capture people's typical writing tasks. We see it as a trade-off: %
It allowed people in the study to start from a given material while requiring them to write a new text (about the article). Starting without any input would likely require a much longer study to experience some aspects of the writing process.%
\changenote{
Third, we decided against a baseline setup where users write text without summarization support. We did this to maximize the writing time spent using our system. However, even without a comparative baseline some of the participants compared our system to their usual writing process (see \cref{sec:results_comparison}).
}

Based on these experiences, we conclude that the writing task to analyze given argumentative texts was a suitable choice. We decided for a controlled study to gain insights into participant's thinking while using the tool, but we encourage future work to also consider in-the-wild studies to assess other writing setups. Related, we expect our system to work better for writing longer texts. Users' feedback mentioned long articles or theses as examples. %
Besides studying longer use, it could also be interesting to compare use for argumentative, informative, and creative texts.

\subsection{AI Role Perception Through Design}
Our first design had a \textit{Replace in Text} button in each card that replaced the corresponding paragraph with the summary. As observations and think-aloud showed, this partly made people perceive summaries as \textit{text suggestions} that should be included in the text, rather than annotations. %
While the use of summaries in one's own text is not inherently wrong, this clearly indicated a miscommunication of the intended role of the AI in our design. We thus changed \textit{Replace in Text} to \textit{Copy to Clipboard} in study 2. This suggests a more ``passive'' role of the summary, which could be put into the text with a dedicated further action (paste). Functionally, it is also more flexible, since users can copy it anywhere they like. 
Observations and think-aloud in study 2 indicate that this indeed shifted perception away from ``text suggestions''. %
This contributes a concrete example of how an arguably small design choice might evoke prior mental models around AI features (e.g. of known text suggestion features) that shape the perceived role of the AI. %

In a broader view, we can position our work in the recent framework for modeling interaction in co-creative AI systems by \citet{Rezwana2022}: Their 2022 survey of 92 co-creative AI systems revealed three predominant interaction models: In two, the AI generates content, either turn-based or in parallel to the user. In the third, the AI (also) evaluates the user's creation, in a turn based manner. With this paper, we contribute to the literature an exploration of another interaction model: The AI generates parallel content that empowers users in their self-evaluation. With this new direction, we also respond to the survey's conclusion ``that the space of possibilities is underutilized''~\cite{Rezwana2022}. Considering their insights, future explorations of said space could address implicit human-AI communication (e.g. gaze-informed summaries).

\subsection{Designing Human-AI Co-creative Systems from Existing Cognitive Strategies}

In their work on envisioning less ``obvious'' NLP applications, Yang et al.~\cite{Yang2019} highlight that ``[a]uthors are inherently better than algorithms at comprehending their unfinished writing''. Consequently, they propose to reframe the author-AI relationship in terms of other NLP problems (conversation, retrieval/search, question answering). In this way, they used NLP concepts to inform interaction design, whereas we used human writing concepts (i.e. reverse outlining) to do so. We consider both approaches as complementary directions to explore and ``[...] expand this narrow intersection between what is [of] value to users and what can be built''~\cite{Yang2019}.

Dissecting this, our approach starts with a strategy in the target domain (reverse outlining in writing) and asks how it might inspire writing support with NLP. Concretely, the reverse outlining strategy guided and thus greatly facilitated our decision-making for initial design choices (e.g. layout, main interactions) -- even if it was not our goal to ```force'' writers to apply exactly this strategy when using the resulting system. Crucially, it also usefully constrained the role of the AI: Summarize only, no interpretation, no ``idea'' generation. 
However, getting the main features and their UI right required more than this starting guide, as the first study and design iteration showed (e.g. redesigned zoom levels, change from ``replace text'' to ``copy to clipboard''). 
Overall, we recommend this approach based on our experiences here. Looking ahead, existing human strategies (and their respective treatments in writing research) might serve the text interaction community in a role akin to generative theory (cf.~\cite{Beaudouin-Lafon2021}). Our approach provides a template: We can examine other human writing strategies to inspire new NLP tools for writers. %

\section{Conclusion}\label{sec:conclusion}

We have investigated how writers can be supported in reflecting and revising their text. We have found that automatic AI-generated summaries help writers by providing a continuously updated overview and an external perspective during the writing process. Participants used these insights to check and revise the scope and structure of their text draft. This work demonstrates writing support beyond the current focus on text generation and correction. More generally, we envision future AI-powered writing tools to offer a mix of direct text edits and indirect reflection support for writers. We encourage future work to explore how other adaptive margin annotations may help writers in the text drafting process and also how non-textual annotations can be integrated. To facilitate such research in co-creative writing tools, we release the prototype and further material on the project website:

\url{https://osf.io/v6zfn}

\begin{acks}
We thank Christina Schneegass and Lukas Mecke for their feedback on the manuscript. This project is funded by the Bavarian State Ministry of Science and the Arts and coordinated by the Bavarian Research Institute for Digital Transformation (bidt).
\end{acks}

\bibliographystyle{ACM-Reference-Format}
\bibliography{bibliography}


\begin{thebibliography}{58}

%%% the Web address.

\ifx \showCODEN    \undefined \def \showCODEN     #1{\unskip}     \fi
\ifx \showDOI      \undefined \def \showDOI       #1{#1}\fi
\ifx \showISBNx    \undefined \def \showISBNx     #1{\unskip}     \fi
\ifx \showISBNxiii \undefined \def \showISBNxiii  #1{\unskip}     \fi
\ifx \showISSN     \undefined \def \showISSN      #1{\unskip}     \fi
\ifx \showLCCN     \undefined \def \showLCCN      #1{\unskip}     \fi
\ifx \shownote     \undefined \def \shownote      #1{#1}          \fi
\ifx \showarticletitle \undefined \def \showarticletitle #1{#1}   \fi
\ifx \showURL      \undefined \def \showURL       {\relax}        \fi
% invisible to TeX
\providecommand\bibfield[2]{#2}
\providecommand\bibinfo[2]{#2}
\providecommand\natexlab[1]{#1}
\providecommand\showeprint[2][]{arXiv:#2}

\bibitem[\protect\citeauthoryear{Arif, Kim, Stuerzlinger, Lee, and
  Mazalek}{Arif et~al\mbox{.}}{2016}]%
        {Arif2016}
\bibfield{author}{\bibinfo{person}{Ahmed~Sabbir Arif}, \bibinfo{person}{Sunjun
  Kim}, \bibinfo{person}{Wolfgang Stuerzlinger}, \bibinfo{person}{Geehyuk Lee},
  {and} \bibinfo{person}{Ali Mazalek}.} \bibinfo{year}{2016}\natexlab{}.
\newblock \showarticletitle{Evaluation of a Smart-Restorable Backspace
  Technique to Facilitate Text Entry Error Correction}. In
  \bibinfo{booktitle}{\emph{Proceedings of the 2016 CHI Conference on Human
  Factors in Computing Systems}} (San Jose, California, USA)
  \emph{(\bibinfo{series}{CHI '16})}. \bibinfo{publisher}{Association for
  Computing Machinery}, \bibinfo{address}{New York, NY, USA},
  \bibinfo{pages}{5151–5162}.
\newblock
\showISBNx{9781450333627}
\urldef\tempurl%
\url{https://doi.org/10.1145/2858036.2858407}
\showDOI{\tempurl}


\bibitem[\protect\citeauthoryear{Arnold, Volzer, and Madrid}{Arnold
  et~al\mbox{.}}{2021}]%
        {arnold_generative_2021}
\bibfield{author}{\bibinfo{person}{Kenneth~C Arnold}, \bibinfo{person}{April~M
  Volzer}, {and} \bibinfo{person}{Noah~G Madrid}.}
  \bibinfo{year}{2021}\natexlab{}.
\newblock \showarticletitle{Generative {Models} can {Help} {Writers} without
  {Writing} for {Them}}.
\newblock \bibinfo{journal}{\emph{2nd Workshop on Human-AI Co-Creation with
  Generative Models - HAI-GEN 2021}} (\bibinfo{year}{2021}),
  \bibinfo{pages}{8}.
\newblock


\bibitem[\protect\citeauthoryear{Beaudouin-Lafon, B\o{}dker, and
  Mackay}{Beaudouin-Lafon et~al\mbox{.}}{2021}]%
        {Beaudouin-Lafon2021}
\bibfield{author}{\bibinfo{person}{Michel Beaudouin-Lafon},
  \bibinfo{person}{Susanne B\o{}dker}, {and} \bibinfo{person}{Wendy~E.
  Mackay}.} \bibinfo{year}{2021}\natexlab{}.
\newblock \showarticletitle{Generative Theories of Interaction}.
\newblock \bibinfo{journal}{\emph{ACM Trans. Comput.-Hum. Interact.}}
  \bibinfo{volume}{28}, \bibinfo{number}{6}, Article \bibinfo{articleno}{45}
  (\bibinfo{date}{nov} \bibinfo{year}{2021}), \bibinfo{numpages}{54}~pages.
\newblock
\showISSN{1073-0516}
\urldef\tempurl%
\url{https://doi.org/10.1145/3468505}
\showDOI{\tempurl}


\bibitem[\protect\citeauthoryear{Buschek, Z\"{u}rn, and Eiband}{Buschek
  et~al\mbox{.}}{2021}]%
        {Buschek2021chi}
\bibfield{author}{\bibinfo{person}{Daniel Buschek}, \bibinfo{person}{Martin
  Z\"{u}rn}, {and} \bibinfo{person}{Malin Eiband}.}
  \bibinfo{year}{2021}\natexlab{}.
\newblock \showarticletitle{The Impact of Multiple Parallel Phrase Suggestions
  on Email Input and Composition Behaviour of Native and Non-Native English
  Writers}. In \bibinfo{booktitle}{\emph{Proceedings of the 2021 CHI Conference
  on Human Factors in Computing Systems}} (Yokohama, Japan)
  \emph{(\bibinfo{series}{CHI '21})}. \bibinfo{publisher}{Association for
  Computing Machinery}, \bibinfo{address}{New York, NY, USA}, Article
  \bibinfo{articleno}{732}, \bibinfo{numpages}{13}~pages.
\newblock
\showISBNx{9781450380966}
\urldef\tempurl%
\url{https://doi.org/10.1145/3411764.3445372}
\showDOI{\tempurl}


\bibitem[\protect\citeauthoryear{Ceylan and Mihalcea}{Ceylan and
  Mihalcea}{2009}]%
        {gelbukh_decomposition_2009}
\bibfield{author}{\bibinfo{person}{Hakan Ceylan} {and} \bibinfo{person}{Rada
  Mihalcea}.} \bibinfo{year}{2009}\natexlab{}.
\newblock \showarticletitle{The {Decomposition} of {Human}-{Written} {Book}
  {Summaries}}.
\newblock In \bibinfo{booktitle}{\emph{Computational {Linguistics} and
  {Intelligent} {Text} {Processing}}},
  \bibfield{editor}{\bibinfo{person}{Alexander Gelbukh}} (Ed.).
  Vol.~\bibinfo{volume}{5449}. \bibinfo{publisher}{Springer Berlin Heidelberg},
  \bibinfo{address}{Berlin, Heidelberg}, \bibinfo{pages}{582--593}.
\newblock
\showISBNx{978-3-642-00381-3 978-3-642-00382-0}
\urldef\tempurl%
\url{https://doi.org/10.1007/978-3-642-00382-0_47}
\showDOI{\tempurl}
\newblock
\shownote{Series Title: Lecture Notes in Computer Science.}


\bibitem[\protect\citeauthoryear{Chen, Lee, Bansal, Cao, Zhang, Lu, Tsay, Wang,
  Dai, Chen, Sohn, and Wu}{Chen et~al\mbox{.}}{2019}]%
        {Chen2019}
\bibfield{author}{\bibinfo{person}{Mia~Xu Chen}, \bibinfo{person}{Benjamin~N.
  Lee}, \bibinfo{person}{Gagan Bansal}, \bibinfo{person}{Yuan Cao},
  \bibinfo{person}{Shuyuan Zhang}, \bibinfo{person}{Justin Lu},
  \bibinfo{person}{Jackie Tsay}, \bibinfo{person}{Yinan Wang},
  \bibinfo{person}{Andrew~M. Dai}, \bibinfo{person}{Zhifeng Chen},
  \bibinfo{person}{Timothy Sohn}, {and} \bibinfo{person}{Yonghui Wu}.}
  \bibinfo{year}{2019}\natexlab{}.
\newblock \showarticletitle{Gmail Smart Compose: Real-Time Assisted Writing}.
  In \bibinfo{booktitle}{\emph{Proceedings of the 25th ACM SIGKDD International
  Conference on Knowledge Discovery \& Data Mining}} (Anchorage, AK, USA)
  \emph{(\bibinfo{series}{KDD ’19})}. \bibinfo{publisher}{Association for
  Computing Machinery}, \bibinfo{address}{New York, NY, USA},
  \bibinfo{pages}{2287–2295}.
\newblock
\showISBNx{9781450362016}
\urldef\tempurl%
\url{https://doi.org/10.1145/3292500.3330723}
\showDOI{\tempurl}


\bibitem[\protect\citeauthoryear{Chung, Kim, Yoo, Lee, Adar, and Chang}{Chung
  et~al\mbox{.}}{2022}]%
        {chung_talebrush_2022}
\bibfield{author}{\bibinfo{person}{John Joon~Young Chung},
  \bibinfo{person}{Wooseok Kim}, \bibinfo{person}{Kang~Min Yoo},
  \bibinfo{person}{Hwaran Lee}, \bibinfo{person}{Eytan Adar}, {and}
  \bibinfo{person}{Minsuk Chang}.} \bibinfo{year}{2022}\natexlab{}.
\newblock \showarticletitle{{TaleBrush}: {Sketching} {Stories} with
  {Generative} {Pretrained} {Language} {Models}}. In
  \bibinfo{booktitle}{\emph{Proceedings of the 2022 {CHI} {Conference} on
  {Human} {Factors} in {Computing} {Systems}}}, Vol.~\bibinfo{volume}{CHI'22}.
  \bibinfo{publisher}{Association for Computing Machinery},
  \bibinfo{address}{New Orleans, LA, USA}, \bibinfo{pages}{19}.
\newblock


\bibitem[\protect\citeauthoryear{Clark, Ross, Tan, Ji, and Smith}{Clark
  et~al\mbox{.}}{2018}]%
        {clark2018}
\bibfield{author}{\bibinfo{person}{Elizabeth Clark},
  \bibinfo{person}{Anne~Spencer Ross}, \bibinfo{person}{Chenhao Tan},
  \bibinfo{person}{Yangfeng Ji}, {and} \bibinfo{person}{Noah~A. Smith}.}
  \bibinfo{year}{2018}\natexlab{}.
\newblock \showarticletitle{Creative Writing with a Machine in the Loop: Case
  Studies on Slogans and Stories}. In \bibinfo{booktitle}{\emph{23rd
  International Conference on Intelligent User Interfaces}} (Tokyo, Japan)
  \emph{(\bibinfo{series}{IUI '18})}. \bibinfo{publisher}{Association for
  Computing Machinery}, \bibinfo{address}{New York, NY, USA},
  \bibinfo{pages}{329–340}.
\newblock
\showISBNx{9781450349451}
\urldef\tempurl%
\url{https://doi.org/10.1145/3172944.3172983}
\showDOI{\tempurl}


\bibitem[\protect\citeauthoryear{Corbin}{Corbin}{1990}]%
        {corbin1990basics}
\bibfield{author}{\bibinfo{person}{Juliet~M Corbin}.}
  \bibinfo{year}{1990}\natexlab{}.
\newblock \bibinfo{booktitle}{\emph{Basics of qualitative research: Grounded
  theory procedures and techniques}}.
\newblock \bibinfo{publisher}{Sage}.
\newblock


\bibitem[\protect\citeauthoryear{Cresswell}{Cresswell}{2000}]%
        {Cresswell2000}
\bibfield{author}{\bibinfo{person}{Andy Cresswell}.}
  \bibinfo{year}{2000}\natexlab{}.
\newblock \showarticletitle{Self-monitoring in student writing: developing
  learner responsibility}.
\newblock \bibinfo{journal}{\emph{ELT Journal}} \bibinfo{volume}{54},
  \bibinfo{number}{3} (\bibinfo{date}{July} \bibinfo{year}{2000}),
  \bibinfo{pages}{235--244}.
\newblock
\showISSN{0951-0893}
\urldef\tempurl%
\url{https://doi.org/10.1093/elt/54.3.235}
\showDOI{\tempurl}


\bibitem[\protect\citeauthoryear{Cui, Zhu, Zhang, Schwartz, Wobbrock, and
  Bi}{Cui et~al\mbox{.}}{2020}]%
        {Cui2020}
\bibfield{author}{\bibinfo{person}{Wenzhe Cui}, \bibinfo{person}{Suwen Zhu},
  \bibinfo{person}{Mingrui~Ray Zhang}, \bibinfo{person}{H.~Andrew Schwartz},
  \bibinfo{person}{Jacob~O. Wobbrock}, {and} \bibinfo{person}{Xiaojun Bi}.}
  \bibinfo{year}{2020}\natexlab{}.
\newblock \bibinfo{booktitle}{\emph{JustCorrect: Intelligent Post Hoc Text
  Correction Techniques on Smartphones}}.
\newblock \bibinfo{publisher}{Association for Computing Machinery},
  \bibinfo{address}{New York, NY, USA}, \bibinfo{pages}{487–499}.
\newblock
\showISBNx{9781450375146}
\urldef\tempurl%
\url{https://doi.org/10.1145/3379337.3415857}
\showURL{%
\tempurl}


\bibitem[\protect\citeauthoryear{Flower and Hayes}{Flower and Hayes}{1981}]%
        {flower1981cognitive}
\bibfield{author}{\bibinfo{person}{Linda Flower} {and} \bibinfo{person}{John~R
  Hayes}.} \bibinfo{year}{1981}\natexlab{}.
\newblock \showarticletitle{A cognitive process theory of writing}.
\newblock \bibinfo{journal}{\emph{College composition and communication}}
  \bibinfo{volume}{32}, \bibinfo{number}{4} (\bibinfo{year}{1981}),
  \bibinfo{pages}{365--387}.
\newblock


\bibitem[\protect\citeauthoryear{Gero and Chilton}{Gero and Chilton}{2019}]%
        {Gero2019chi}
\bibfield{author}{\bibinfo{person}{Katy~Ilonka Gero} {and}
  \bibinfo{person}{Lydia~B. Chilton}.} \bibinfo{year}{2019}\natexlab{}.
\newblock \showarticletitle{Metaphoria: An Algorithmic Companion for Metaphor
  Creation}. In \bibinfo{booktitle}{\emph{Proceedings of the 2019 CHI
  Conference on Human Factors in Computing Systems}} (Glasgow, Scotland Uk)
  \emph{(\bibinfo{series}{CHI ’19})}. \bibinfo{publisher}{Association for
  Computing Machinery}, \bibinfo{address}{New York, NY, USA},
  \bibinfo{pages}{1–12}.
\newblock
\showISBNx{9781450359702}
\urldef\tempurl%
\url{https://doi.org/10.1145/3290605.3300526}
\showDOI{\tempurl}


\bibitem[\protect\citeauthoryear{Ghazvininejad, Shi, Priyadarshi, and
  Knight}{Ghazvininejad et~al\mbox{.}}{2017}]%
        {Hafez2017}
\bibfield{author}{\bibinfo{person}{Marjan Ghazvininejad}, \bibinfo{person}{Xing
  Shi}, \bibinfo{person}{Jay Priyadarshi}, {and} \bibinfo{person}{Kevin
  Knight}.} \bibinfo{year}{2017}\natexlab{}.
\newblock \showarticletitle{{H}afez: an Interactive Poetry Generation System}.
  In \bibinfo{booktitle}{\emph{Proceedings of {ACL} 2017, System
  Demonstrations}}. \bibinfo{publisher}{Association for Computational
  Linguistics}, \bibinfo{address}{Vancouver, Canada}, \bibinfo{pages}{43--48}.
\newblock
\urldef\tempurl%
\url{https://www.aclweb.org/anthology/P17-4008}
\showURL{%
\tempurl}


\bibitem[\protect\citeauthoryear{Hamburger}{Hamburger}{2013}]%
        {hamburger_outlining_2013}
\bibfield{author}{\bibinfo{person}{Aaron Hamburger}.}
  \bibinfo{year}{2013}\natexlab{}.
\newblock \bibinfo{title}{Outlining in {Reverse}}.
\newblock
\newblock
\urldef\tempurl%
\url{https://opinionator.blogs.nytimes.com/2013/01/21/outlining-in-reverse/}
\showURL{%
\tempurl}
\newblock
\shownote{Cad: 1 Section: Opinion.}


\bibitem[\protect\citeauthoryear{Han, Renom, Mackay, and Beaudouin-Lafon}{Han
  et~al\mbox{.}}{2020}]%
        {Han2020}
\bibfield{author}{\bibinfo{person}{Han~L. Han}, \bibinfo{person}{Miguel~A.
  Renom}, \bibinfo{person}{Wendy~E. Mackay}, {and} \bibinfo{person}{Michel
  Beaudouin-Lafon}.} \bibinfo{year}{2020}\natexlab{}.
\newblock \bibinfo{booktitle}{\emph{Textlets: Supporting Constraints and
  Consistency in Text Documents}}.
\newblock \bibinfo{publisher}{Association for Computing Machinery},
  \bibinfo{address}{New York, NY, USA}, \bibinfo{pages}{1–13}.
\newblock
\showISBNx{9781450367080}
\urldef\tempurl%
\url{https://doi.org/10.1145/3313831.3376804}
\showURL{%
\tempurl}


\bibitem[\protect\citeauthoryear{Huang, Schoop, Ha, and Canny}{Huang
  et~al\mbox{.}}{2020}]%
        {Huang2020}
\bibfield{author}{\bibinfo{person}{Forrest Huang}, \bibinfo{person}{Eldon
  Schoop}, \bibinfo{person}{David Ha}, {and} \bibinfo{person}{John Canny}.}
  \bibinfo{year}{2020}\natexlab{}.
\newblock \showarticletitle{Scones: Towards Conversational Authoring of
  Sketches}. In \bibinfo{booktitle}{\emph{Proceedings of the 25th International
  Conference on Intelligent User Interfaces}} (Cagliari, Italy)
  \emph{(\bibinfo{series}{IUI '20})}. \bibinfo{publisher}{Association for
  Computing Machinery}, \bibinfo{address}{New York, NY, USA},
  \bibinfo{pages}{313–323}.
\newblock
\showISBNx{9781450371186}
\urldef\tempurl%
\url{https://doi.org/10.1145/3377325.3377485}
\showDOI{\tempurl}


\bibitem[\protect\citeauthoryear{Kannan, Kurach, Ravi, Kaufmann, Tomkins,
  Miklos, Corrado, Lukacs, Ganea, Young, and Ramavajjala}{Kannan
  et~al\mbox{.}}{2016}]%
        {Kannan2016}
\bibfield{author}{\bibinfo{person}{Anjuli Kannan}, \bibinfo{person}{Karol
  Kurach}, \bibinfo{person}{Sujith Ravi}, \bibinfo{person}{Tobias Kaufmann},
  \bibinfo{person}{Andrew Tomkins}, \bibinfo{person}{Balint Miklos},
  \bibinfo{person}{Greg Corrado}, \bibinfo{person}{Laszlo Lukacs},
  \bibinfo{person}{Marina Ganea}, \bibinfo{person}{Peter Young}, {and}
  \bibinfo{person}{Vivek Ramavajjala}.} \bibinfo{year}{2016}\natexlab{}.
\newblock \showarticletitle{Smart Reply: Automated Response Suggestion for
  Email}. In \bibinfo{booktitle}{\emph{Proceedings of the 22nd ACM SIGKDD
  International Conference on Knowledge Discovery and Data Mining}} (San
  Francisco, California, USA) \emph{(\bibinfo{series}{KDD ’16})}.
  \bibinfo{publisher}{Association for Computing Machinery},
  \bibinfo{address}{New York, NY, USA}, \bibinfo{pages}{955–964}.
\newblock
\showISBNx{9781450342322}
\urldef\tempurl%
\url{https://doi.org/10.1145/2939672.2939801}
\showDOI{\tempurl}


\bibitem[\protect\citeauthoryear{King}{King}{2012}]%
        {king_reverse_2012}
\bibfield{author}{\bibinfo{person}{Cynthia~L. King}.}
  \bibinfo{year}{2012}\natexlab{}.
\newblock \showarticletitle{Reverse {Outlining}: {A} {Method} for {Effective}
  {Revision} of {Document} {Structure}}.
\newblock \bibinfo{journal}{\emph{IEEE Transactions on Professional
  Communication}} \bibinfo{volume}{55}, \bibinfo{number}{3}
  (\bibinfo{date}{Sept.} \bibinfo{year}{2012}), \bibinfo{pages}{254--261}.
\newblock
\showISSN{1558-1500}
\urldef\tempurl%
\url{https://doi.org/10.1109/TPC.2012.2207838}
\showDOI{\tempurl}
\newblock
\shownote{Conference Name: IEEE Transactions on Professional Communication.}


\bibitem[\protect\citeauthoryear{Kryscinski, Keskar, McCann, Xiong, and
  Socher}{Kryscinski et~al\mbox{.}}{2019}]%
        {Kryscinski2019}
\bibfield{author}{\bibinfo{person}{Wojciech Kryscinski},
  \bibinfo{person}{Nitish~Shirish Keskar}, \bibinfo{person}{Bryan McCann},
  \bibinfo{person}{Caiming Xiong}, {and} \bibinfo{person}{Richard Socher}.}
  \bibinfo{year}{2019}\natexlab{}.
\newblock \showarticletitle{Neural Text Summarization: A Critical Evaluation}.
  In \bibinfo{booktitle}{\emph{Proceedings of the 2019 Conference on Empirical
  Methods in Natural Language Processing and the 9th International Joint
  Conference on Natural Language Processing (EMNLP-IJCNLP)}}.
  \bibinfo{publisher}{Association for Computational Linguistics},
  \bibinfo{address}{Hong Kong, China}, \bibinfo{pages}{540--551}.
\newblock
\urldef\tempurl%
\url{https://doi.org/10.18653/v1/D19-1051}
\showDOI{\tempurl}


\bibitem[\protect\citeauthoryear{Lab}{Lab}{2021}]%
        {lab_reverse_2021}
\bibfield{author}{\bibinfo{person}{Purdue~Writing Lab}.}
  \bibinfo{year}{2021}\natexlab{}.
\newblock \bibinfo{title}{Reverse {Outlining} // {Purdue} {Writing} {Lab}}.
\newblock
\newblock
\urldef\tempurl%
\url{https://owl.purdue.edu/owl/general_writing/the_writing_process/reverse_outlining.html}
\showURL{%
\tempurl}


\bibitem[\protect\citeauthoryear{Lee, Liang, and Yang}{Lee
  et~al\mbox{.}}{2022}]%
        {lee_coauthor_2022}
\bibfield{author}{\bibinfo{person}{Mina Lee}, \bibinfo{person}{Percy Liang},
  {and} \bibinfo{person}{Qian Yang}.} \bibinfo{year}{2022}\natexlab{}.
\newblock \showarticletitle{{CoAuthor}: {Designing} a {Human}-{AI}
  {Collaborative} {Writing} {Dataset} for {Exploring} {Language} {Model}
  {Capabilities}}. In \bibinfo{booktitle}{\emph{Proceedings of the 2022 {CHI}
  {Conference} on {Human} {Factors} in {Computing} {Systems}}}
  \emph{(\bibinfo{series}{{CHI}'22})}. \bibinfo{publisher}{Association for
  Computing Machinery}, \bibinfo{address}{New Orleans, LA, USA}.
\newblock
\urldef\tempurl%
\url{https://doi.org/10.1145/3491102.3502030}
\showDOI{\tempurl}


\bibitem[\protect\citeauthoryear{Leiva}{Leiva}{2018}]%
        {Leiva2018}
\bibfield{author}{\bibinfo{person}{Luis~A. Leiva}.}
  \bibinfo{year}{2018}\natexlab{}.
\newblock \showarticletitle{Responsive text summarization}.
\newblock \bibinfo{journal}{\emph{Inform. Process. Lett.}}
  \bibinfo{volume}{130} (\bibinfo{year}{2018}), \bibinfo{pages}{52--57}.
\newblock
\showISSN{0020-0190}
\urldef\tempurl%
\url{https://doi.org/10.1016/j.ipl.2017.10.007}
\showDOI{\tempurl}


\bibitem[\protect\citeauthoryear{LeVan and King}{LeVan and King}{2017}]%
        {levan2017self}
\bibfield{author}{\bibinfo{person}{Karen~Sheriff LeVan} {and}
  \bibinfo{person}{Marissa~E King}.} \bibinfo{year}{2017}\natexlab{}.
\newblock \showarticletitle{Self-Annotation as a Course Practice}.
\newblock \bibinfo{journal}{\emph{Teaching English in the Two Year College}}
  \bibinfo{volume}{44}, \bibinfo{number}{3} (\bibinfo{year}{2017}),
  \bibinfo{pages}{289}.
\newblock


\bibitem[\protect\citeauthoryear{Li, Chen, Tung, and Chilton}{Li
  et~al\mbox{.}}{2021}]%
        {Li2021uist}
\bibfield{author}{\bibinfo{person}{Daniel Li}, \bibinfo{person}{Thomas Chen},
  \bibinfo{person}{Albert Tung}, {and} \bibinfo{person}{Lydia~B Chilton}.}
  \bibinfo{year}{2021}\natexlab{}.
\newblock \showarticletitle{Hierarchical Summarization for Longform Spoken
  Dialog}. In \bibinfo{booktitle}{\emph{The 34th Annual ACM Symposium on User
  Interface Software and Technology}} (Virtual Event, USA)
  \emph{(\bibinfo{series}{UIST '21})}. \bibinfo{publisher}{Association for
  Computing Machinery}, \bibinfo{address}{New York, NY, USA},
  \bibinfo{pages}{582–597}.
\newblock
\showISBNx{9781450386357}
\urldef\tempurl%
\url{https://doi.org/10.1145/3472749.3474771}
\showDOI{\tempurl}


\bibitem[\protect\citeauthoryear{Li, Sarcar, Kim, and Ren}{Li
  et~al\mbox{.}}{2020}]%
        {Li2020}
\bibfield{author}{\bibinfo{person}{Yang Li}, \bibinfo{person}{Sayan Sarcar},
  \bibinfo{person}{Sunjun Kim}, {and} \bibinfo{person}{Xiangshi Ren}.}
  \bibinfo{year}{2020}\natexlab{}.
\newblock \bibinfo{booktitle}{\emph{Swap: A Replacement-Based Text Revision
  Technique for Mobile Devices}}.
\newblock \bibinfo{publisher}{Association for Computing Machinery},
  \bibinfo{address}{New York, NY, USA}, \bibinfo{pages}{1–12}.
\newblock
\showISBNx{9781450367080}
\urldef\tempurl%
\url{https://doi.org/10.1145/3313831.3376217}
\showURL{%
\tempurl}


\bibitem[\protect\citeauthoryear{Macdonald, Frase, Gingrich, and
  Keenan}{Macdonald et~al\mbox{.}}{1982}]%
        {macdonald1982writer}
\bibfield{author}{\bibinfo{person}{Nina Macdonald}, \bibinfo{person}{Lawrence
  Frase}, \bibinfo{person}{P Gingrich}, {and} \bibinfo{person}{Stacey Keenan}.}
  \bibinfo{year}{1982}\natexlab{}.
\newblock \showarticletitle{The writer's workbench: Computer aids for text
  analysis}.
\newblock \bibinfo{journal}{\emph{IEEE Transactions on Communications}}
  \bibinfo{volume}{30}, \bibinfo{number}{1} (\bibinfo{year}{1982}),
  \bibinfo{pages}{105--110}.
\newblock


\bibitem[\protect\citeauthoryear{Messuri}{Messuri}{2016a}]%
        {messuri_revision_2016}
\bibfield{author}{\bibinfo{person}{Kristin Messuri}.}
  \bibinfo{year}{2016}\natexlab{a}.
\newblock \showarticletitle{Revision {Strategies}}.
\newblock \bibinfo{journal}{\emph{The Southwest Respiratory and Critical Care
  Chronicles}} \bibinfo{volume}{4}, \bibinfo{number}{14} (\bibinfo{date}{April}
  \bibinfo{year}{2016}), \bibinfo{pages}{46--48}.
\newblock
\showISSN{2325-9205}
\urldef\tempurl%
\url{https://pulmonarychronicles.com/index.php/pulmonarychronicles/article/view/263}
\showURL{%
\tempurl}
\newblock
\shownote{Number: 14.}


\bibitem[\protect\citeauthoryear{Messuri}{Messuri}{2016b}]%
        {messuri_writing_2016}
\bibfield{author}{\bibinfo{person}{Kristin Messuri}.}
  \bibinfo{year}{2016}\natexlab{b}.
\newblock \showarticletitle{Writing {Effective} {Paragraphs}}.
\newblock \bibinfo{journal}{\emph{The Southwest Respiratory and Critical Care
  Chronicles}} \bibinfo{volume}{4}, \bibinfo{number}{15} (\bibinfo{date}{July}
  \bibinfo{year}{2016}), \bibinfo{pages}{86--88}.
\newblock
\showISSN{2325-9205}
\urldef\tempurl%
\url{https://pulmonarychronicles.com/index.php/pulmonarychronicles/article/view/290}
\showURL{%
\tempurl}
\newblock
\shownote{Number: 15.}


\bibitem[\protect\citeauthoryear{Mihalcea and Tarau}{Mihalcea and
  Tarau}{2004}]%
        {Mihalcea2004textrank}
\bibfield{author}{\bibinfo{person}{Rada Mihalcea} {and} \bibinfo{person}{Paul
  Tarau}.} \bibinfo{year}{2004}\natexlab{}.
\newblock \showarticletitle{Textrank: Bringing order into text}. In
  \bibinfo{booktitle}{\emph{Proceedings of the 2004 conference on empirical
  methods in natural language processing}}. \bibinfo{pages}{404--411}.
\newblock


\bibitem[\protect\citeauthoryear{Muller and Kogan}{Muller and Kogan}{2012}]%
        {muller_grounded_2012}
\bibfield{author}{\bibinfo{person}{Michael~J. Muller} {and}
  \bibinfo{person}{Sandra Kogan}.} \bibinfo{year}{2012}\natexlab{}.
\newblock \showarticletitle{Grounded {Theory} {Method} in {Human}-{Computer}
  {Interaction} and {Computer}-{Supported} {Cooperative} {Work}}.
\newblock In \bibinfo{booktitle}{\emph{The {Human}–{Computer} {Interaction}
  {Handbook}} (\bibinfo{edition}{3} ed.)}. \bibinfo{publisher}{CRC Press}.
\newblock
\showISBNx{978-0-429-10397-1}
\newblock
\shownote{Num Pages: 21.}


\bibitem[\protect\citeauthoryear{of~the Center for Excellence~in
  Writing}{of~the Center for Excellence~in Writing}{2007}]%
        {uni_tennessee_writing_strategies_2007}
\bibfield{author}{\bibinfo{person}{Graduate Writing~Center of~the Center for
  Excellence~in Writing}.} \bibinfo{year}{2007}\natexlab{}.
\newblock \bibinfo{title}{Strategies for {Drafting} \& {Revising} {Academic}
  {Writing}}.
\newblock
\newblock
\urldef\tempurl%
\url{https://www.tnstate.edu/write/documents/DraftingRevisingEves2007.pdf}
\showURL{%
\tempurl}


\bibitem[\protect\citeauthoryear{of~Wisconsin-Madison}{of~Wisconsin-Madison}{2021}]%
        {university_of_wisconsin-madison_reverse_2021}
\bibfield{author}{\bibinfo{person}{University of Wisconsin-Madison}.}
  \bibinfo{year}{2021}\natexlab{}.
\newblock \bibinfo{title}{Reverse {Outlines}: {A} {Writer}’s {Technique} for
  {Examining} {Organization}}.
\newblock
\newblock
\urldef\tempurl%
\url{https://writing.wisc.edu/wp-content/uploads/sites/535/2018/07/reverseoutlines_uwmadison_writingcenter_aug2012.pdf}
\showURL{%
\tempurl}


\bibitem[\protect\citeauthoryear{Park}{Park}{2008}]%
        {park_reverse_2008}
\bibfield{author}{\bibinfo{person}{Sam Park}.} \bibinfo{year}{2008}\natexlab{}.
\newblock \bibinfo{title}{Reverse {Outlining} {Worksheet} {\textbar} {Student}
  {Learning} {Center}}.
\newblock
\newblock
\urldef\tempurl%
\url{https://slc.berkeley.edu/writing-worksheets-and-other-writing-resources/reverse-outlining-worksheet}
\showURL{%
\tempurl}


\bibitem[\protect\citeauthoryear{Radev, Hovy, and McKeown}{Radev
  et~al\mbox{.}}{2002}]%
        {radev_introduction_2002}
\bibfield{author}{\bibinfo{person}{Dragomir~R. Radev}, \bibinfo{person}{Eduard
  Hovy}, {and} \bibinfo{person}{Kathleen McKeown}.}
  \bibinfo{year}{2002}\natexlab{}.
\newblock \showarticletitle{Introduction to the {Special} {Issue} on
  {Summarization}}.
\newblock \bibinfo{journal}{\emph{Computational Linguistics}}
  \bibinfo{volume}{28}, \bibinfo{number}{4} (\bibinfo{date}{Dec.}
  \bibinfo{year}{2002}), \bibinfo{pages}{399--408}.
\newblock
\showISSN{0891-2017, 1530-9312}
\urldef\tempurl%
\url{https://doi.org/10.1162/089120102762671927}
\showDOI{\tempurl}


\bibitem[\protect\citeauthoryear{Raffel, Shazeer, Roberts, Lee, Narang, Matena,
  Zhou, Li, and Liu}{Raffel et~al\mbox{.}}{2020}]%
        {Raffel2020}
\bibfield{author}{\bibinfo{person}{Colin Raffel}, \bibinfo{person}{Noam
  Shazeer}, \bibinfo{person}{Adam Roberts}, \bibinfo{person}{Katherine Lee},
  \bibinfo{person}{Sharan Narang}, \bibinfo{person}{Michael Matena},
  \bibinfo{person}{Yanqi Zhou}, \bibinfo{person}{Wei Li}, {and}
  \bibinfo{person}{Peter~J. Liu}.} \bibinfo{year}{2020}\natexlab{}.
\newblock \showarticletitle{Exploring the {Limits} of {Transfer} {Learning}
  with a {Unified} {Text}-to-{Text} {Transformer}}.
\newblock \bibinfo{journal}{\emph{arXiv:1910.10683 [cs, stat]}}
  (\bibinfo{date}{July} \bibinfo{year}{2020}).
\newblock
\urldef\tempurl%
\url{http://arxiv.org/abs/1910.10683}
\showURL{%
\tempurl}
\newblock
\shownote{arXiv: 1910.10683.}


\bibitem[\protect\citeauthoryear{Rezwana and Maher}{Rezwana and Maher}{2022}]%
        {Rezwana2022}
\bibfield{author}{\bibinfo{person}{Jeba Rezwana} {and}
  \bibinfo{person}{Mary~Lou Maher}.} \bibinfo{year}{2022}\natexlab{}.
\newblock \showarticletitle{Designing Creative AI Partners with COFI: A
  Framework for Modeling Interaction in Human-AI Co-Creative Systems}.
\newblock \bibinfo{journal}{\emph{ACM Trans. Comput.-Hum. Interact.}}
  (\bibinfo{date}{feb} \bibinfo{year}{2022}).
\newblock
\showISSN{1073-0516}
\urldef\tempurl%
\url{https://doi.org/10.1145/3519026}
\showDOI{\tempurl}
\newblock
\shownote{Just Accepted.}


\bibitem[\protect\citeauthoryear{Roemmele and Gordon}{Roemmele and
  Gordon}{2015}]%
        {roemmele2015creative}
\bibfield{author}{\bibinfo{person}{Melissa Roemmele} {and}
  \bibinfo{person}{Andrew~S Gordon}.} \bibinfo{year}{2015}\natexlab{}.
\newblock \showarticletitle{Creative help: A story writing assistant}. In
  \bibinfo{booktitle}{\emph{International Conference on Interactive Digital
  Storytelling}}. Springer, \bibinfo{pages}{81--92}.
\newblock


\bibitem[\protect\citeauthoryear{Saltz}{Saltz}{1998}]%
        {saltz_harvard_1998}
\bibfield{author}{\bibinfo{person}{Laura Saltz}.}
  \bibinfo{year}{1998}\natexlab{}.
\newblock \bibinfo{title}{Harvard {College} {Writing} {Center} - {Revising} the
  {Draft}}.
\newblock
\newblock
\urldef\tempurl%
\url{https://writingcenter.fas.harvard.edu/pages/revising-draft}
\showURL{%
\tempurl}


\bibitem[\protect\citeauthoryear{Schmitt and Buschek}{Schmitt and
  Buschek}{2021}]%
        {Schmitt2021}
\bibfield{author}{\bibinfo{person}{Oliver Schmitt} {and}
  \bibinfo{person}{Daniel Buschek}.} \bibinfo{year}{2021}\natexlab{}.
\newblock \showarticletitle{CharacterChat: Supporting the Creation of Fictional
  Characters through Conversation and Progressive Manifestation with a
  Chatbot}. In \bibinfo{booktitle}{\emph{Creativity and Cognition}} (Virtual
  Event, Italy) \emph{(\bibinfo{series}{C\&C '21})}.
  \bibinfo{publisher}{Association for Computing Machinery},
  \bibinfo{address}{New York, NY, USA}, Article \bibinfo{articleno}{10},
  \bibinfo{numpages}{10}~pages.
\newblock
\showISBNx{9781450383769}
\urldef\tempurl%
\url{https://doi.org/10.1145/3450741.3465253}
\showDOI{\tempurl}


\bibitem[\protect\citeauthoryear{See, Liu, and Manning}{See
  et~al\mbox{.}}{2017}]%
        {See2017}
\bibfield{author}{\bibinfo{person}{Abigail See}, \bibinfo{person}{Peter Liu},
  {and} \bibinfo{person}{Christopher Manning}.}
  \bibinfo{year}{2017}\natexlab{}.
\newblock \showarticletitle{Get To The Point: Summarization with
  Pointer-Generator Networks}. In \bibinfo{booktitle}{\emph{Association for
  Computational Linguistics}}.
\newblock
\urldef\tempurl%
\url{https://arxiv.org/abs/1704.04368}
\showURL{%
\tempurl}


\bibitem[\protect\citeauthoryear{Singh, Bernal, Savchenko, and Glassman}{Singh
  et~al\mbox{.}}{2022}]%
        {elephant_tochi2022}
\bibfield{author}{\bibinfo{person}{Nikhil Singh}, \bibinfo{person}{Guillermo
  Bernal}, \bibinfo{person}{Daria Savchenko}, {and} \bibinfo{person}{Elena~L.
  Glassman}.} \bibinfo{year}{2022}\natexlab{}.
\newblock \showarticletitle{Where to Hide a Stolen Elephant: Leaps in Creative
  Writing with Multimodal Machine Intelligence}.
\newblock \bibinfo{journal}{\emph{ACM Trans. Comput.-Hum. Interact.}}
  (\bibinfo{date}{jan} \bibinfo{year}{2022}).
\newblock
\showISSN{1073-0516}
\urldef\tempurl%
\url{https://doi.org/10.1145/3511599}
\showDOI{\tempurl}
\newblock
\shownote{Just Accepted.}


\bibitem[\protect\citeauthoryear{Strobl, Ailhaud, Benetos, Devitt, Kruse,
  Proske, and Rapp}{Strobl et~al\mbox{.}}{2019}]%
        {Strobl2019}
\bibfield{author}{\bibinfo{person}{Carola Strobl}, \bibinfo{person}{Emilie
  Ailhaud}, \bibinfo{person}{Kalliopi Benetos}, \bibinfo{person}{Ann Devitt},
  \bibinfo{person}{Otto Kruse}, \bibinfo{person}{Antje Proske}, {and}
  \bibinfo{person}{Christian Rapp}.} \bibinfo{year}{2019}\natexlab{}.
\newblock \showarticletitle{Digital support for academic writing: A review of
  technologies and pedagogies}.
\newblock \bibinfo{journal}{\emph{Computers \& Education}}
  \bibinfo{volume}{131} (\bibinfo{year}{2019}), \bibinfo{pages}{33--48}.
\newblock
\showISSN{0360-1315}
\urldef\tempurl%
\url{https://doi.org/10.1016/j.compedu.2018.12.005}
\showDOI{\tempurl}


\bibitem[\protect\citeauthoryear{Subramonyam, Seifert, Shah, and
  Adar}{Subramonyam et~al\mbox{.}}{2020}]%
        {Subramonyam2020}
\bibfield{author}{\bibinfo{person}{Hariharan Subramonyam},
  \bibinfo{person}{Colleen Seifert}, \bibinfo{person}{Priti Shah}, {and}
  \bibinfo{person}{Eytan Adar}.} \bibinfo{year}{2020}\natexlab{}.
\newblock \bibinfo{booktitle}{\emph{TexSketch: Active Diagramming through
  Pen-and-Ink Annotations}}.
\newblock \bibinfo{publisher}{Association for Computing Machinery},
  \bibinfo{address}{New York, NY, USA}, \bibinfo{pages}{1–13}.
\newblock
\showISBNx{9781450367080}
\urldef\tempurl%
\url{https://doi.org/10.1145/3313831.3376155}
\showURL{%
\tempurl}


\bibitem[\protect\citeauthoryear{Swanson and Gordon}{Swanson and
  Gordon}{2008}]%
        {swanson2008say}
\bibfield{author}{\bibinfo{person}{Reid Swanson} {and}
  \bibinfo{person}{Andrew~S Gordon}.} \bibinfo{year}{2008}\natexlab{}.
\newblock \showarticletitle{Say anything: A massively collaborative open domain
  story writing companion}. In \bibinfo{booktitle}{\emph{Joint International
  Conference on Interactive Digital Storytelling}}. Springer,
  \bibinfo{pages}{32--40}.
\newblock


\bibitem[\protect\citeauthoryear{Tambwekar, Dhuliawala, Martin, Mehta,
  Harrison, and Riedl}{Tambwekar et~al\mbox{.}}{2018}]%
        {Tambwekar2018controllable}
\bibfield{author}{\bibinfo{person}{Pradyumna Tambwekar},
  \bibinfo{person}{Murtaza Dhuliawala}, \bibinfo{person}{Lara~J. Martin},
  \bibinfo{person}{Animesh Mehta}, \bibinfo{person}{Brent Harrison}, {and}
  \bibinfo{person}{Mark~O. Riedl}.} \bibinfo{year}{2018}\natexlab{}.
\newblock \bibinfo{title}{Controllable Neural Story Plot Generation via
  Reinforcement Learning}.
\newblock
\newblock
\showeprint[arxiv]{1809.10736}~[cs.CL]


\bibitem[\protect\citeauthoryear{ter Hoeve, Sim, Nouri, Fourney, de~Rijke, and
  White}{ter Hoeve et~al\mbox{.}}{2020}]%
        {terHoeve2020}
\bibfield{author}{\bibinfo{person}{Maartje ter Hoeve}, \bibinfo{person}{Robert
  Sim}, \bibinfo{person}{Elnaz Nouri}, \bibinfo{person}{Adam Fourney},
  \bibinfo{person}{Maarten de Rijke}, {and} \bibinfo{person}{Ryen~W. White}.}
  \bibinfo{year}{2020}\natexlab{}.
\newblock \showarticletitle{Conversations with Documents: An Exploration of
  Document-Centered Assistance}. In \bibinfo{booktitle}{\emph{Proceedings of
  the 2020 Conference on Human Information Interaction and Retrieval}}
  (Vancouver BC, Canada) \emph{(\bibinfo{series}{CHIIR '20})}.
  \bibinfo{publisher}{Association for Computing Machinery},
  \bibinfo{address}{New York, NY, USA}, \bibinfo{pages}{43–52}.
\newblock
\showISBNx{9781450368926}
\urldef\tempurl%
\url{https://doi.org/10.1145/3343413.3377971}
\showDOI{\tempurl}


\bibitem[\protect\citeauthoryear{Tully}{Tully}{2019}]%
        {tully_reverse_2019}
\bibfield{author}{\bibinfo{person}{L.~Danielle Tully}.}
  \bibinfo{year}{2019}\natexlab{}.
\newblock \showarticletitle{Reverse {Outlines}: {Fueling} {Revision} \&
  {Preparing} for {Writing} {Conferences}}.
\newblock \bibinfo{journal}{\emph{The Second Draft}} \bibinfo{volume}{32},
  \bibinfo{number}{2} (\bibinfo{year}{2019}), \bibinfo{pages}{6}.
\newblock
\urldef\tempurl%
\url{https://ssrn.com/abstract=3465807}
\showURL{%
\tempurl}


\bibitem[\protect\citeauthoryear{University}{University}{2021}]%
        {duke_university_revising_2021}
\bibfield{author}{\bibinfo{person}{Duke University}.}
  \bibinfo{year}{2021}\natexlab{}.
\newblock \bibinfo{title}{Revising {Process} {\textbar} {Thompson} {Writing}
  {Program}}.
\newblock
\newblock
\urldef\tempurl%
\url{https://twp.duke.edu/sites/twp.duke.edu/files/file-attachments/reverse-outline.original.pdf}
\showURL{%
\tempurl}


\bibitem[\protect\citeauthoryear{Vertanen, Dunlop, Clawson, Kristensson, and
  Arif}{Vertanen et~al\mbox{.}}{2016}]%
        {Vertanen2016}
\bibfield{author}{\bibinfo{person}{Keith Vertanen}, \bibinfo{person}{Mark
  Dunlop}, \bibinfo{person}{James Clawson}, \bibinfo{person}{Per~Ola
  Kristensson}, {and} \bibinfo{person}{Ahmed~Sabbir Arif}.}
  \bibinfo{year}{2016}\natexlab{}.
\newblock \showarticletitle{Inviscid Text Entry and Beyond}. In
  \bibinfo{booktitle}{\emph{Proceedings of the 2016 CHI Conference Extended
  Abstracts on Human Factors in Computing Systems}} (San Jose, California, USA)
  \emph{(\bibinfo{series}{CHI EA '16})}. \bibinfo{publisher}{Association for
  Computing Machinery}, \bibinfo{address}{New York, NY, USA},
  \bibinfo{pages}{3469–3476}.
\newblock
\showISBNx{9781450340823}
\urldef\tempurl%
\url{https://doi.org/10.1145/2851581.2856472}
\showDOI{\tempurl}


\bibitem[\protect\citeauthoryear{Vertanen, Montague, Dunlop, Arif, Bi, and
  Azenkot}{Vertanen et~al\mbox{.}}{2017}]%
        {Vertanen2017}
\bibfield{author}{\bibinfo{person}{Keith Vertanen}, \bibinfo{person}{Kyle
  Montague}, \bibinfo{person}{Mark Dunlop}, \bibinfo{person}{Ahmed~Sabbir
  Arif}, \bibinfo{person}{Xiaojun Bi}, {and} \bibinfo{person}{Shiri Azenkot}.}
  \bibinfo{year}{2017}\natexlab{}.
\newblock \showarticletitle{Ubiquitous Text Interaction}. In
  \bibinfo{booktitle}{\emph{Proceedings of the 2017 CHI Conference Extended
  Abstracts on Human Factors in Computing Systems}} (Denver, Colorado, USA)
  \emph{(\bibinfo{series}{CHI EA '17})}. \bibinfo{publisher}{Association for
  Computing Machinery}, \bibinfo{address}{New York, NY, USA},
  \bibinfo{pages}{566–573}.
\newblock
\showISBNx{9781450346566}
\urldef\tempurl%
\url{https://doi.org/10.1145/3027063.3027066}
\showDOI{\tempurl}


\bibitem[\protect\citeauthoryear{Wang, Li, Zhou, Chen, Grossman, and Li}{Wang
  et~al\mbox{.}}{2021}]%
        {wang2021screen2words}
\bibfield{author}{\bibinfo{person}{Bryan Wang}, \bibinfo{person}{Gang Li},
  \bibinfo{person}{Xin Zhou}, \bibinfo{person}{Zhourong Chen},
  \bibinfo{person}{Tovi Grossman}, {and} \bibinfo{person}{Yang Li}.}
  \bibinfo{year}{2021}\natexlab{}.
\newblock \showarticletitle{Screen2words: Automatic mobile UI summarization
  with multimodal learning}. In \bibinfo{booktitle}{\emph{The 34th Annual ACM
  Symposium on User Interface Software and Technology}}.
  \bibinfo{pages}{498--510}.
\newblock


\bibitem[\protect\citeauthoryear{Waters and Schneider}{Waters and
  Schneider}{2009}]%
        {waters_metacognition_2009}
\bibfield{author}{\bibinfo{person}{Harriet~Salatas Waters} {and}
  \bibinfo{person}{Wolfgang Schneider}.} \bibinfo{year}{2009}\natexlab{}.
\newblock \bibinfo{booktitle}{\emph{Metacognition, {Strategy} {Use}, and
  {Instruction}}}.
\newblock \bibinfo{publisher}{Guilford Press}.
\newblock
\showISBNx{978-1-60623-384-9}


\bibitem[\protect\citeauthoryear{Yang, Zhou, Zhang, and Li}{Yang
  et~al\mbox{.}}{2022}]%
        {yang_ai_2022}
\bibfield{author}{\bibinfo{person}{Daijin Yang}, \bibinfo{person}{Yanpeng
  Zhou}, \bibinfo{person}{Zhiyuan Zhang}, {and} \bibinfo{person}{Toby Jia-Jun
  Li}.} \bibinfo{year}{2022}\natexlab{}.
\newblock \showarticletitle{{AI} as an {Active} {Writer}: {Interaction}
  strategies with generated text in human-{AI} collaborative fiction writing}.
  In \bibinfo{booktitle}{\emph{{IUI} 2022 {Workshop} on {Human}-{AI}
  {Co}-{Creation} with {Generative} {Models} ({HAI}-{GEN} 2022)}}.
  \bibinfo{pages}{10}.
\newblock


\bibitem[\protect\citeauthoryear{Yang, Cranshaw, Amershi, Iqbal, and
  Teevan}{Yang et~al\mbox{.}}{2019}]%
        {Yang2019}
\bibfield{author}{\bibinfo{person}{Qian Yang}, \bibinfo{person}{Justin
  Cranshaw}, \bibinfo{person}{Saleema Amershi}, \bibinfo{person}{Shamsi~T.
  Iqbal}, {and} \bibinfo{person}{Jaime Teevan}.}
  \bibinfo{year}{2019}\natexlab{}.
\newblock \showarticletitle{Sketching NLP: A Case Study of Exploring the Right
  Things To Design with Language Intelligence}. In
  \bibinfo{booktitle}{\emph{Proceedings of the 2019 CHI Conference on Human
  Factors in Computing Systems}} (Glasgow, Scotland Uk)
  \emph{(\bibinfo{series}{CHI '19})}. \bibinfo{publisher}{Association for
  Computing Machinery}, \bibinfo{address}{New York, NY, USA},
  \bibinfo{pages}{1–12}.
\newblock
\showISBNx{9781450359702}
\urldef\tempurl%
\url{https://doi.org/10.1145/3290605.3300415}
\showDOI{\tempurl}


\bibitem[\protect\citeauthoryear{Yayl{\i}}{Yayl{\i}}{2012}]%
        {yayli2012tracing}
\bibfield{author}{\bibinfo{person}{Demet Yayl{\i}}.}
  \bibinfo{year}{2012}\natexlab{}.
\newblock \showarticletitle{Tracing the benefits of self annotation in
  genre-based writing}.
\newblock \bibinfo{journal}{\emph{The Journal of Language Learning and
  Teaching}} \bibinfo{volume}{2}, \bibinfo{number}{1} (\bibinfo{year}{2012}),
  \bibinfo{pages}{45--58}.
\newblock


\bibitem[\protect\citeauthoryear{Yuan, Coenen, Reif, and Ippolito}{Yuan
  et~al\mbox{.}}{2022}]%
        {yuan_wordcraft_2022}
\bibfield{author}{\bibinfo{person}{Ann Yuan}, \bibinfo{person}{Andy Coenen},
  \bibinfo{person}{Emily Reif}, {and} \bibinfo{person}{Daphne Ippolito}.}
  \bibinfo{year}{2022}\natexlab{}.
\newblock \showarticletitle{Wordcraft: {Story} {Writing} {With} {Large}
  {Language} {Models}}. In \bibinfo{booktitle}{\emph{27th {International}
  {Conference} on {Intelligent} {User} {Interfaces}}}
  \emph{(\bibinfo{series}{{IUI} '22})}. \bibinfo{publisher}{Association for
  Computing Machinery}, \bibinfo{address}{New York, NY, USA},
  \bibinfo{pages}{841--852}.
\newblock
\showISBNx{978-1-4503-9144-3}
\urldef\tempurl%
\url{https://doi.org/10.1145/3490099.3511105}
\showDOI{\tempurl}


\bibitem[\protect\citeauthoryear{Zhang, Wen, and Wobbrock}{Zhang
  et~al\mbox{.}}{2019}]%
        {Zhang2019}
\bibfield{author}{\bibinfo{person}{Mingrui~Ray Zhang}, \bibinfo{person}{He
  Wen}, {and} \bibinfo{person}{Jacob~O. Wobbrock}.}
  \bibinfo{year}{2019}\natexlab{}.
\newblock \showarticletitle{Type, Then Correct: Intelligent Text Correction
  Techniques for Mobile Text Entry Using Neural Networks}. In
  \bibinfo{booktitle}{\emph{Proceedings of the 32nd Annual ACM Symposium on
  User Interface Software and Technology}} (New Orleans, LA, USA)
  \emph{(\bibinfo{series}{UIST '19})}. \bibinfo{publisher}{Association for
  Computing Machinery}, \bibinfo{address}{New York, NY, USA},
  \bibinfo{pages}{843–855}.
\newblock
\showISBNx{9781450368162}
\urldef\tempurl%
\url{https://doi.org/10.1145/3332165.3347924}
\showDOI{\tempurl}


\end{thebibliography}

\appendix

\section{Additional Model Evaluation}

\changenote{
To assess the quality of the summaries displayed in the study, we report on an additional evaluation. Ideally, we would evaluate summary quality after each interaction by comparing the automatically created summary in each card to a human-created summary for that paragraph. This is not practical because the text and thus summaries change all the time during interaction.
}

\begin{table}[]
\begin{tabular}{@{}lll@{}}
\toprule
Average Rouge Score  & Central sentences & Abstractive summaries \\ \midrule
Rouge-1 R & 0.5503            & 0.2344                \\
Rouge-1 P & 0.5316            & 0.2588                \\
Rouge-1 F & 0.5355            & 0.2302                \\
Rouge-2 R  & 0.4886            & 0.0705                \\
Rouge-2 P & 0.4930            & 0.0895                \\
Rouge-2 F & 0.4897            & 0.0723                \\
Rouge-l R & 0.5422            & 0.2127                \\
Rouge-l P & 0.5266            & 0.2349                \\
Rouge-l F & 0.5295            & 0.2087                \\ \bottomrule
\end{tabular}
\caption{Results from the ROUGE analysis of the T5 model against human selected central sentences and human written (abstractive) summaries.}
\label{tab:rouge}
\end{table}

\changenote{
Instead, we thus decided to assess summarization quality on the \textit{given articles}. To do so, we recruited three people who wrote abstractive summaries for each paragraph of the articles. They also selected what they thought is the most central sentence per paragraph.
}

\changenote{
We used this data to compute ROUGE scores (\cref{tab:rouge}). The mean ROUGE-L F-score is ca. 0.2 against human summaries, and ca. 0.5 against human-selected sentences. This is generally comparable to the ROUGE scores reported on the benchmark dataset in the T5 paper~\cite{Raffel2020}. 
}

\changenote{
This indicates that the T5 model performed similarly to its published benchmark when used on the domains that people wrote about in our study. %
}

\changenote{
However, note that the dataset and reference generation method are different. Thus, we mainly report these values with the goal of providing a point of comparison for future work that employs AI summaries in systems and use-cases similar to ours here.
}

\changenote{
Finally, we also compared the sentences selected by our central sentence method against the sentences selected by the three human annotators. They matched (i.e. same sentence chosen by system and annotators) in 48\,\% of the cases.
}

\end{document}